# Theory of Supercoupling, Squeezing Wave Energy, and Field Confinement in Narrow Channels and Tight Bends Using $\varepsilon$-Near-Zero Metamaterials

*Mário Silveirinha*[(1,2)] *and Nader Engheta*[(1)*]

*(1) University of Pennsylvania, Department of Electrical and Systems Engineering, Philadelphia, PA, U.S.A., engheta@ee.upenn.edu*
*(2) Universidade de Coimbra, Department of Electrical Engineering – Instituto de Telecomunicações, Portugal, mario.silveirinha@co.it.pt*

**Abstract**

In this work, we investigate the detailed theory of the supercoupling, anomalous tunneling effect, and field confinement originally identified in [M. Silveirinha, N. Engheta, Phys. Rev. Lett. 97, 157403, (2006)], where we demonstrated the possibility of using materials with permittivity $\varepsilon$-near zero to drastically improve the transmission of electromagnetic energy through a narrow irregular channel with very subwavelength transverse cross-section. Here, we present additional physical insights, describe new applications of the tunneling effect in relevant waveguide scenarios (e.g., the "perfect" or "super" waveguide coupling), study the effect of metal losses in the metallic walls, and the possibility of using $\varepsilon$-near zero materials to confine energy in a subwavelength cavity with gigantic field enhancement. In addition, we systematically study the propagation of electromagnetic waves through narrow channels filled with anisotropic $\varepsilon$-near zero materials. It is demonstrated that these materials may have interesting potentials, and that for some particular geometries the reflectivity of the channel is independent of the specific dimensions or parameters of $\varepsilon$-near zero transition. We also describe several realistic metamaterial implementations of the studied problems, based on standard metallic waveguides, microstrip line configurations, and wire media.

---

[*] To whom correspondence should be addressed: E-mail: engheta@ee.upenn.edu

# I. Introduction

In recent years there has been a growing interest in the development of new technologies or approaches that potentially allow confining and guiding electromagnetic energy with mode sizes below the diffraction limit. This may have key applications in several fields such as telecommunications (e.g. realization of compact cavities or waveguides [1, 2]), imaging with subwavelength resolution [3, 4], devices with increased storage capacity, delivery and concentration (nanofocusing) of the optical radiation energy on the nanoscale [5, 6, 7], and realization of compact optical resonators [8]. Most of these proposals rely on the excitation of surface plasmon polaritons (SPP) supported by metallic structures with negative permittivity. In our recent work [9], we proposed a different paradigm to break the diffraction limit and squeeze light through channels and bends with subwavelength cross-section. We theoretically demonstrated that if a narrow channel is filled with a $\varepsilon$-near-zero (ENZ) material then, in the lossless limit, its reflectivity only depends on the *volume* of the channel, being independent of its specific geometry and of the transverse cross-section (relatively to the direction of propagation). Moreover, in a counterintuitive way, our results establish that the transmission through the ENZ channel is improved when the transverse cross-section is made more and more narrow. These properties suggest that ENZ materials may have interesting potentials in efficiently transmitting energy through subwavelength regions and effectively providing 'supercoupling" between two ports/waveguides.

In [10], we have further shown that by loading the ENZ material with dielectric or metallic particles it is possible to tailor the magnetic permeability of the material. In this way, by suitably designing the inclusions it is possible to tune the permeability of the ENZ filling material, without changing its electric properties. In particular, it may be feasible to design a matched zero-index metamaterial having both permittivity and permeability near zero, and thus improved transmission characteristics. A remarkable property of such matched zero-index materials is that the way they interact with electromagnetic waves is independent of the granularity of the composite material or of the specific lattice arrangement.

The objective of the present work is to study with more detail the theory of the supercoupling, field confinement, and tunneling effect reported in [9], giving new physical insights of this phenomenon, and describing its emergence and potential applications in new configurations and scenarios. Namely, here we investigate the group velocity of the wave as it travels through the ENZ channel, evaluate the effect of losses in the metallic walls, and demonstrate the possibility of concentrating and confining the electromagnetic fields in a very subwavelength air cavity with enormous electric field enhancement. Moreover, we propose a realistic (metamaterial) emulation of the studied propagation scenarios at microwaves using standard 'empty' metallic waveguides or alternatively using a microstrip line configuration. We report results obtained with a full wave electromagnetic simulator that demonstrate the emergence of a similar tunneling and supercoupling effect in these very realistic, and arguably simple, setups.

In addition, following the ideas of [9], we exploit the possibility of using *anisotropic* ENZ materials to create an analogous tunneling effect. In fact, as referred to in [9], while at infrared and optical frequencies (isotropic) materials with $\varepsilon \approx 0$ may be readily available in nature (e.g. some metals [11], semiconductors [12], or polar dielectrics [13]), at microwaves these materials are not readily accessible (an exception is the electron gas, of which the ionosphere is a well-known example at radiowaves). Nevertheless, these materials may in principle be fabricated as artificial microstructured materials. However, nowadays the fabrication of *isotropic* ENZ materials is still relatively more challenging due to the complexity of the required *isotropic* microstructure of the material. To circumvent this problem, in [9] we suggested using anisotropic ENZ materials − which are comparatively simpler to synthesize − in order to obtain the same tunneling effect. In this work, we further develop these concepts, and develop a detailed theory for the propagation of electromagnetic waves through narrow channels filled with anisotropic ENZ materials. We also discuss the design of wire-medium-based implementations of these anisotropic materials at microwaves.

This paper is organized as follows. In section II we investigate the supercoupling properties of metallic channels filled with isotropic ENZ materials. The effect of metallic and dielectric losses is discussed, as well as possible applications (e.g., for waveguide coupling) and specific features of the tunneling phenomenon. It is also described how to emulate the studied propagation scenarios at microwaves using realistic 3D configurations based on the concept of artificial plasma. In section III, we study the propagation of waves through anisotropic ENZ materials. It is shown that for some configurations the scattering parameters may be independent of the specific geometry of the structure. The design of anisotropic ENZ slabs using wire media is discussed in detail, enlightening new features and characteristics of such metamaterials in propagation scenarios of interest. Finally, in section IV the conclusions are drawn.

The time variation of the electromagnetic fields is assumed of the form $e^{-i\omega t}$, where $\omega$ is the angular frequency and $i = \sqrt{-1}$.

# II. Supercoupling and Squeezing Energy through ENZ Isotropic Channels

In this section, we present new physical insights related with the propagation of electromagnetic waves through metallic channels filled with ENZ isotropic materials. Important aspects such as the group velocity and the effect of losses in the metallic walls are analyzed. The possibility of concentrating the electric field in a small air cavity with gigantic enhancement is suggested.

## A. *Overview and physical insights of the tunneling effect*

Here, we briefly review the main results derived in [9] and describe the emergence of the tunneling effect in several propagation scenarios. In [9] we studied a generic two-dimensional problem (with geometry invariant along the *z*-direction), and we assumed that the polarization of the fields is such that $\mathbf{H} = H_z(x,y)\hat{\mathbf{u}}_z$. We studied the properties of the electromagnetic fields inside a material with $\varepsilon \approx 0$. It was demonstrated that in order that the electric field is finite inside the ENZ material, it is necessary (in the lossless limit) that $H_z = const.$ inside the material. We used this fundamental result to characterize the transmission of energy through a generic ENZ transition in a waveguide scenario. More specifically, we examined a configuration in which two parallel-plate waveguides are interfaced by an ENZ channel of arbitrary shape. We found out that when a transverse electromagnetic mode (TEM) impinges on the ENZ channel the reflection coefficient (for the magnetic field) is given by (in the $\varepsilon \approx 0$ lossless limit and assuming that the walls of the metallic waveguides are perfectly electric conducting (PEC) materials):

$$\rho = \frac{(a_1 - a_2) + ik_0 \mu_{r,p} A_p}{(a_1 + a_2) - ik_0 \mu_{r,p} A_p}, \tag{1}$$

where $k_0 = \omega/c$ is the free-space wave number, $a_1$ and $a_2$ define the spacing between the metallic plates of the input and output waveguides, respectively, $\mu_{r,p}$ is the relative permeability of the ENZ material and $A_p$ is the *area* of the cross-section of the ENZ channel (contained in the *x-y* plane). The transmission coefficient is given by $T = 1 + \rho$. The geometry of the problem is depicted in Fig. 1 for a very specific ENZ channel in which the transition is shaped as an "U", but we underline here that (1) is valid independently of the precise geometry of the ENZ transition. As discussed in [9], Eq. (1) demonstrates that in the case $a_1 = a_2 \equiv a$, it may be possible to squeeze more and more energy through the channel, as the transverse section of the channel (relatively to the direction of propagation) is made more and more tight, i.e. as $A_p / a$ is made increasingly small. Such effect was demonstrated in [9] for the case of a waveguide with an 180º-bend.

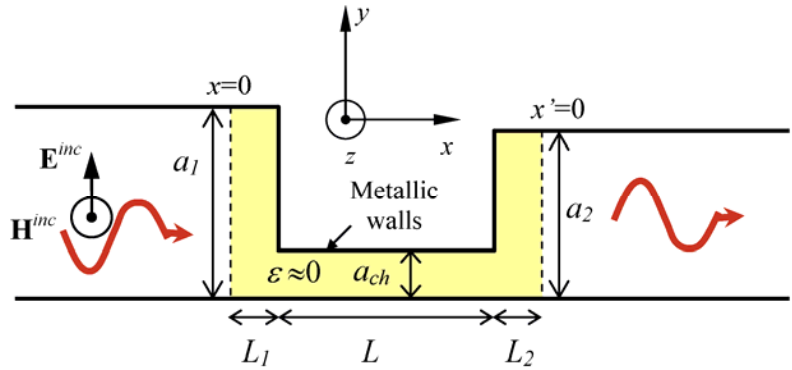

**Fig. 1**. (Color online) Geometry of the two-dimensional problem: two parallel-plate metallic waveguides are interfaced by a U-shaped channel filled with an ENZ material. The incident wave is the fundamental TEM mode. The structure is uniform along the *z*-direction.

Next we further explore some alternative possibilities, namely we present results of full wave simulations computed with CST Microwave Studio$^{TM}$ [14], which demonstrate the emergence of the same tunneling effect in the geometry of Fig. 1 ("U-shaped" transition channel). We will assume that $a_1 = a_2 \equiv a$ since this situation favors the transmission of energy through the channel. It is clear that for such U-shaped channel the area of the cross-section is $A_p = (L_1 + L_2)a + La_{ch}$. Thus, for $a$ and $L$ arbitrarily fixed it is evident that (in the lossless limit) we can make $A_p/a$ arbitrarily small (and consequently make the reflection coefficient approach zero), by reducing more and more the transverse section of the channel $a_{ch}$, and by making the transition regions $L_1$ and $L_2$ more and more thin. Thus, even though the impedance contrast between the free-space region and the ENZ material is infinite the wave may effectively tunnel through the narrow channel with high transmissivity. Note that the previous discussion holds independently of the electrical size of the channel, i.e. of the value of $k_0 L$.

In practice, in a realistic physical system, such an effect is limited by finite losses in the ENZ material and/or by dielectric breakdown. In Fig. 2 we illustrate the effect of losses in the ENZ material. The simulations were obtained for a structure with $L_1 = L_2 = a_{ch} = 0.1a$, and $L = 1.0a$. The ENZ material is characterized by a Drude-type model with relative permittivity $\varepsilon = 1 - \dfrac{\omega_p^2}{\omega(\omega + i\Gamma)}$, where $\omega_p$ is the plasma frequency and $\Gamma$ is the collision frequency [rad/s]. In the simulations we have taken $\omega_p a/c = \pi/2$. Note that at $\omega = \omega_p$ the permittivity of the channel is given by $\varepsilon \approx i\Gamma/\omega_p$. In this simulation the effect of losses in the metallic walls was neglected and will be discussed later in the paper.

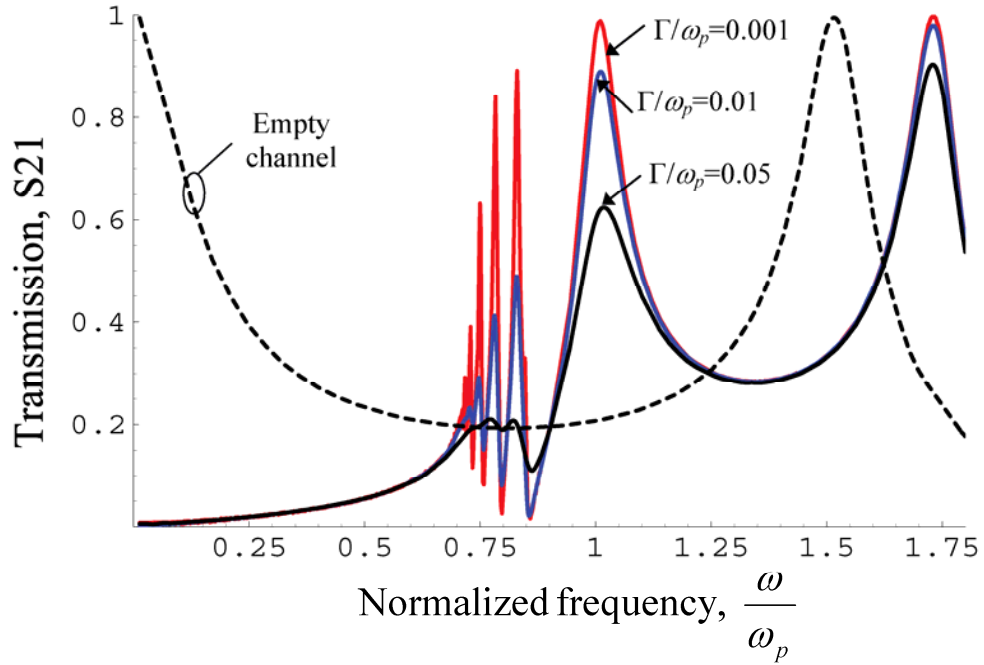

**Fig. 2**. (Color online) Amplitude of the transmission coefficient (S21-parameter) as a function of normalized frequency for the U-shaped ENZ transition depicted in Fig. 1 and different values of the losses $\Gamma/\omega_p$. The dashed line represents the transmission coefficient when the U-shaped transition is filled with air (empty channel).

The results of Fig. 2 confirm that at $\omega = \omega_p$ the wave does, in fact, tunnel through the narrow channel, especially when dielectric losses are small, and that the transmission is still quite significant for moderate losses. In Fig. 2, we also show (dashed line) the transmission coefficient when the U-shaped channel is empty (i.e. filled with air). It is seen that at $\omega = \omega_p$ the wave is unable to propagate around the obstacle and is strongly reflected at the interface. It is worth noting that the transmissivity of the unfilled channel can be quite significant around $\omega/\omega_p \approx 1.5$. This happens due to a geometrical resonance characteristic of the U-shaped geometry, i.e. for certain very specific frequencies related with the very precise values of $L_1$, $L_2$, $L$ and $a_{ch}$, it may be possible to squeeze energy through the U-shaped channel, even though it is filled with air (Fabry-Perot-type transmission). However such an effect is conceptually very different from the effect that can be obtained using an ENZ material. In fact, for a channel filled with an ENZ material, the energy can in principle be squeezed through the channel independently of its specific geometry (e.g. the exact electrical length of the channel). Moreover, (1) predicts that provided $A_p/a$ is kept small the transmitted power is nearly unchanged, even if the precise shape of the channel is radically modified. Thus, in that regard, ENZ materials are indeed unique solutions to squeeze energy through obstructed paths because they effectively create a *zero-order* resonance (electrical length of the channel is zero) that enhances the transmission of energy, independent of the exact geometry of the transition. We also note that below the plasma frequency (in particular for $\text{Re}\{\varepsilon\} \approx -1$, i.e. around $\omega/\omega_p \approx 0.7$) the transmission can also be greatly enhanced due to the excitation of 'quasi-static' localized resonances (local plasmon resonance) characteristic of systems with objects with negative permittivity [15]. These resonances are further analyzed in Appendix A.

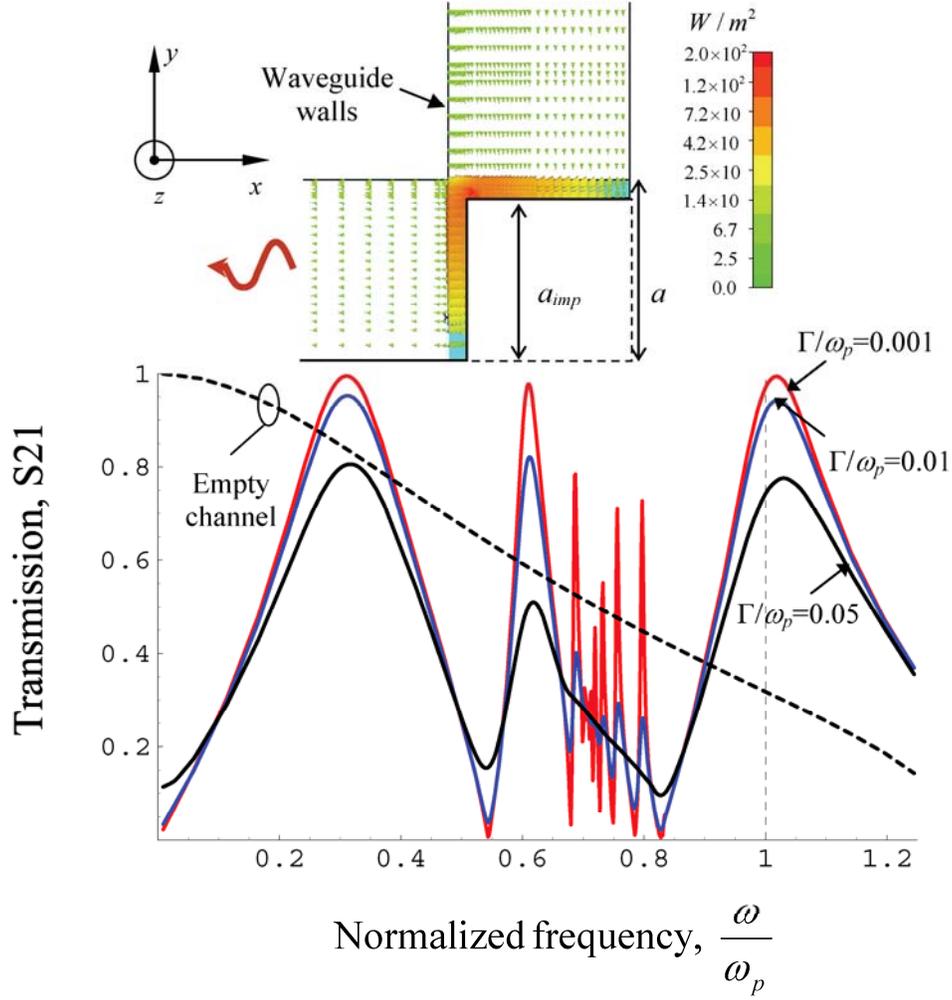

**Fig. 3**. (Color online) Upper panel: Real part of the Poynting vector lines in a waveguide with a 90º-bend filled with an ENZ material at $\omega = \omega_p$ and negligible losses. Lower panel: Amplitude of the transmission coefficient (S21-parameter) as a function of normalized frequency and different values of $\Gamma/\omega_p$. The dashed line represents the transmission coefficient when the 90º-bend is filled with air (empty channel).

It is also interesting to characterize the behavior of the ENZ material in a waveguide configuration with a 90º-bend. The geometry is illustrated in the upper panel of Fig. 3. The incoming wave propagates along the negative *y*-direction. The distance between the parallel metallic plates is *a* (in both the input and output waveguides). The transition between the two waveguides is filled with two thin perpendicular (and connected) ENZ layers with thickness $a - a_{imp}$. In the simulations we assumed that $a_{imp} = 0.9a$, and that the plasma frequency of the ENZ material is such that $\omega_p a / c = 3\pi / 4$. In the lower panel of Fig. 3 we plot the $S_{21}$-parameter (transmission coefficient) as a function of frequency. It is seen that around $\omega = \omega_p$ the wave transmission is greatly enhanced as compared to the case in which the transition is unfilled (empty channel). The upper panel of Fig. 3 depicts the real part of Poynting vector lines, clearly showing that the ENZ material forces the Poynting vector lines to bend and follow the path defined by the shape of the ENZ transition.

At this point it is interesting to discuss why ENZ materials may, in fact, help enhancing transmission through narrow openings. Consider again the U-shaped geometry of Fig. 1. It is reasonable to suppose that the electric field $E_y$ is essentially constant across each transverse section of the channel with height $a_{ch}$ (i.e., to a first approximation $E_y$ is constant at each $x$=const. cut with $L_1 < x < L + L_1$). Thus, using Faraday's law and the fact that in the $\varepsilon = 0$ limit the magnetic field is constant inside the ENZ channel [9] ($H_z = H_z^{int} = const.$), it is evident that for $0 < d < L$ $E_y(d + L_1) = E_y(L_1) + i\omega\mu_0\mu_{r,p}H_z^{int}d$, with $\mu_{r,p}$ being the relative permeability of the ENZ region, or in other words $E_y$ varies linearly along the channel. Thus, defining the transverse wave impedance as $Z_T = E_y / H_z$, we find that:

$$\frac{Z_L}{\eta_0} - \frac{Z_{in}}{\eta_0} \equiv \frac{\delta Z_T}{\eta_0} \approx +i\mu_{r,p}k_0 L, \tag{2}$$

where $\eta_0 = \sqrt{\mu_0/\varepsilon_0}$ is the free-space impedance, and $Z_{in}$ and $Z_L$ are the transverse wave impedances at $x = L_1$ and $x = L + L_1$ (see panel $a$ of Fig. 4). Hence, the wave impedance variation $\delta Z_T$ between the planes $x = L_1$ and $x = L + L_1$ is proportional to the electrical length of the narrow channel.

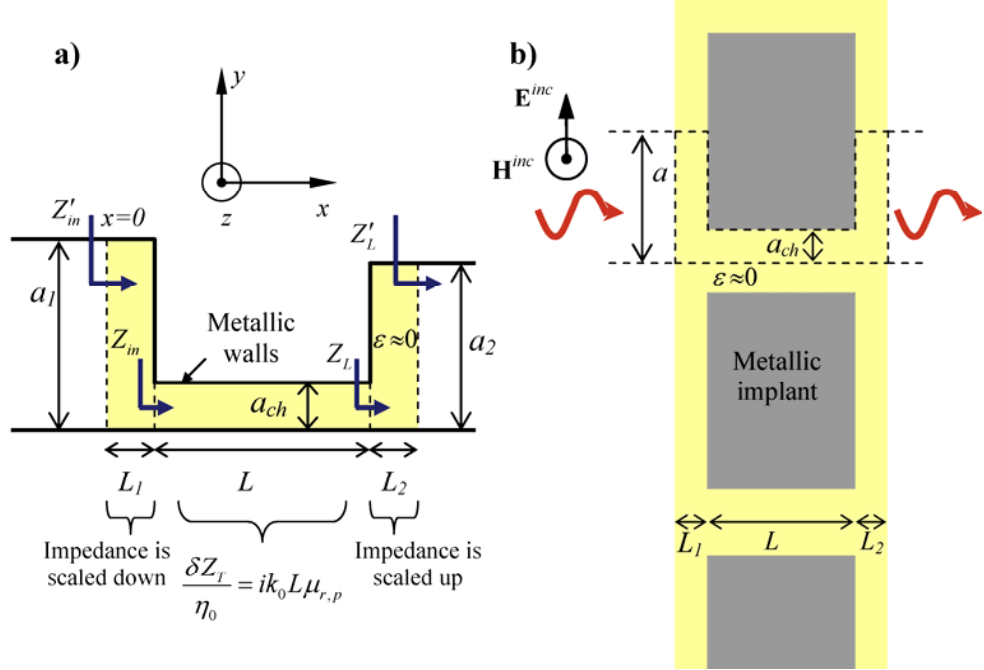

**Fig. 4**. (Color online) Panel (a): Illustration of the wave impedance transformations inside the U-shaped channel. Panel (b): An ENZ slab periodically loaded with metallic implants is illuminated by a plane wave that propagates along the normal direction.

Let $Z'_{in}$ and $Z'_L$ be the transverse wave impedances at the input and output planes ($x = 0$ and $x = L_1 + L + L_2$, respectively), as shown in Fig. 4. Note that in the $\varepsilon = 0$ limit these impedances can be defined unambiguously because the wave in the air regions is a superimposition of TEM waves [9] (in particular, $Z'_L = \eta_0$). It is also evident that in order that the incoming wave can tunnel through the ENZ material it is necessary that $Z'_{in} \approx Z'_L$. However, Eq. (2) shows that the transverse impedance may vary appreciably inside the narrow channel. So, why is it possible to ensure good transmission, even though $\frac{\delta Z_T}{\eta_0}$ may be very large? To find the relation between $Z_{in}$ and $Z'_{in}$, we use again Faraday's law and the relation $H_z = H_z^{int} = const.$. Supposing that $k_0 L_1 \ll 1$, we easily find that $Z'_{in} \approx \frac{a_{ch}}{a_1} Z_{in}$. Similarly, it can be verified that $Z_L \approx \frac{a_2}{a_{ch}} Z'_L$. Thus, the ENZ transition with thickness $L_2$ scales up the load impedance by a factor of $a_2/a_{ch}$, and consequently in general $Z_L/\eta_0$ is very large. Moreover, by making the channel more and more narrow we can always achieve the condition $|\delta Z_T| \ll |Z_L|$, (even though $\delta Z_T/\eta_0$ may also be large), and in this way ensure that $Z_{in}/Z_L \approx 1$. Finally, the ENZ transition with thickness $L_1$ scales down the impedance $Z_{in}$ by a factor $a_{ch}/a_1$. The previous discussion shows that a series of three wave impedance transformations occur inside the ENZ channel (see panel *a* of Fig. 4): first the output impedance ($Z'_L$) is transformed into a very large impedance $Z_L$; next, $Z_L$ is transformed into $Z_{in} = Z_L - \delta Z_T$; finally, $Z_{in}$ is scaled down to the input impedance $Z'_{in}$. In case $a_1 = a_2$, it is clear that the first and third transformations are the inverse of each other. Thus, in these conditions, the wave may tunnel through the channel only when $Z_{in}/Z_L \approx 1$. As explained before, this condition may be ensured by making the channel sufficiently narrow and consequently $|\delta Z_T| \ll |Z_L|$, i.e. by making $Z_L$ sufficiently large the whole system becomes insensitive to the specific (possibly large) value of $\delta Z_T$. This clarifies why the wave may in fact tunnel through the ENZ channel.

The fact that the transmission may be improved when the transverse cross-section of channel is made tighter, can also be explained using the theory and concepts proposed in our previous work [10], where we showed that by loading a ENZ material with metallic or dielectric inclusions it is possible to tailor its magnetic permeability. In fact, for TEM wave incidence and assuming $a_1 = a_2 \equiv a$, the U-shaped waveguide problem is equivalent to the scattering problem depicted in panel *b* of Fig. 4, which shows an infinite ENZ slab periodically loaded with metallic (PEC) inclusions. The ENZ slab is illuminated by a plane wave that propagates along the normal direction. It can be easily verified that the electromagnetic fields in the configuration of panel *b* of Fig. 4 are unaffected (under the previously referred conditions) if PEC walls are placed along the dashed lines that define the contour of the U-shaped geometry. Following [10], the periodically loaded ENZ slab (in the $\varepsilon \approx 0$ lossless limit) is equivalent to an ideal continuous material with effective relative permittivity $\varepsilon_{eff} = 0$ and relative permeability $\mu_{eff} = 1 - f_V$, where $f_V$ is the volume fraction of the metallic inclusions. Note that because the wavelength inside the ENZ material is extremely large, the incoming the wave is insensitive to the granularity of the composite material and to the specific lattice arrangement [10]. Hence, it is clear that by increasing more and more $f_V$ (i.e. by making the ENZ region more and more narrow) the effective permeability of the equivalent composite becomes more and more near zero, i.e. $\mu_{eff} \to 0$ as $f_V \to 1$. This happens because in the static limit good metals have a diamagnetic response. Hence, it is clear that as $f_V \to 1$ the composite medium behaves almost as a matched zero-index material with both $\varepsilon_{eff} = 0$ and $\mu_{eff} = 0$, and thus the impedance mismatch with free-space becomes less and less significant, and consequently the transmission is increased.

Another fundamental property of ENZ materials that gives a different perspective of the tunneling phenomenon is discussed next. As mentioned before, for the 2D problem under study, $H_z = H_z^{\text{int}} = const.$ inside the ENZ material. This result has a very important implication: since the electric current density over the metallic plates is $\mathbf{J}_c = \hat{\mathbf{v}} \times \mathbf{H}$, where $\hat{\mathbf{v}}$ is the unit normal vector (directed to the ENZ region), it follows that $\mathbf{J}_c = H_z^{\text{int}} \hat{\mathbf{t}}$, where $\hat{\mathbf{t}} = \hat{\mathbf{v}} \times \hat{\mathbf{u}}_z$ is the vector tangent to the metallic surface. Thus, the amplitude of the current density is constant, and independent of the specific shape of the plates. Hence, the ENZ material forces the current to follow the path defined by the shape of the metallic plates, and thus the current injected into each PEC plate (left-hand side interface; see Fig. 1), appears undisturbed on the right-hand side interface. That is, the ENZ material, independently of the specific geometry of the channel, preserves the current along the metallic plates. Note that the current flows in opposite directions along the two PEC plates. From a physical point of view, these effects are easy to understand. Indeed, the density of electric charge on the metallic surfaces is $\sigma_c = \varepsilon \mathbf{E}.\hat{\mathbf{v}}$. Since $\varepsilon \approx 0$ in the ENZ material, it follows that $\sigma_c = 0$, and thus, from the conservation of charge, it follows that the amplitude of $\mathbf{J}_c$ in this 2D-problem must, indeed, be invariant along the ENZ channel. Or in other words, independently of the shape of the PEC surface (which can be rather complex inside the ENZ channel), the current is forced to follow the path of the winding PEC footprint without any phase variation since the density of charge on the PEC surface is necessarily zero, and thus $\dfrac{\partial \mathbf{J}_c}{\partial \hat{\mathbf{t}}} = 0$. This property is clearly the physical foundation of the tunneling phenomenon. These results can be generalized to the three-dimensional case, and this will be reported in a future communication.

## *B. Field concentration and confinement in a subwavelength cavity*

As pointed out in [9], when an electromagnetic wave is squeezed through an ENZ channel the electric field inside the ENZ material may be greatly enhanced. This phenomenon is easy to understand using the principle of conservation of energy. In fact, in the lossless limit, the flux of the real part of the Poynting vector through an arbitrary transverse cross-section of the channel must be invariant. Thus, since the magnetic field is constant inside the ENZ material, this requires that the electric field inside the channel is roughly inversely proportional to the height of the channel. This property is also consistent with the discussion of section II.A, where we showed that for the U-shaped geometry the transverse wave impedances at the $x=0$ and $x=L_1$ planes are such that $Z'_{in} \approx \frac{a_{ch}}{a_1} Z_{in}$ (see panel *a* of Fig. 4). This result implies that the electric field $E_y$ must be such that $E_y\big|_{x=0} \approx \frac{a_{ch}}{a_1} E_y\big|_{x=L_1}$ (since $H_z$ is constant inside the ENZ material). In order to illustrate this property, we computed the amplitude of the electric field inside the ENZ channel using CST Microwave Studio$^{TM}$ [14]. The parameters characteristic of the U-shaped channel are the same as in section II.A. In panel *a* of Fig. 5, we depict the Poynting vector lines inside the U-shaped channel. It is seen that consistently with our previous discussion the real part of the Poynting vector is significantly enhanced inside the narrow channel. In panel *b* of the same figure, we plot the amplitude of the electric field along the line $y=0.5a_{ch}$ for different values of $a_{ch}$. It is seen that inside the ENZ channel (in particular for $0.1 < x/a < 1.1$) the electric field is amplified by a factor of $a/a_{ch}$, confirming the results of our theoretical analysis. Specifically, for $a_{ch}/a = 0.1$ the field is enhanced about 10 times inside the ENZ material. Note that even when moderate material losses are taken into account ($\Gamma/\omega_p = 0.05$), the amplification factor may be as large as 6.

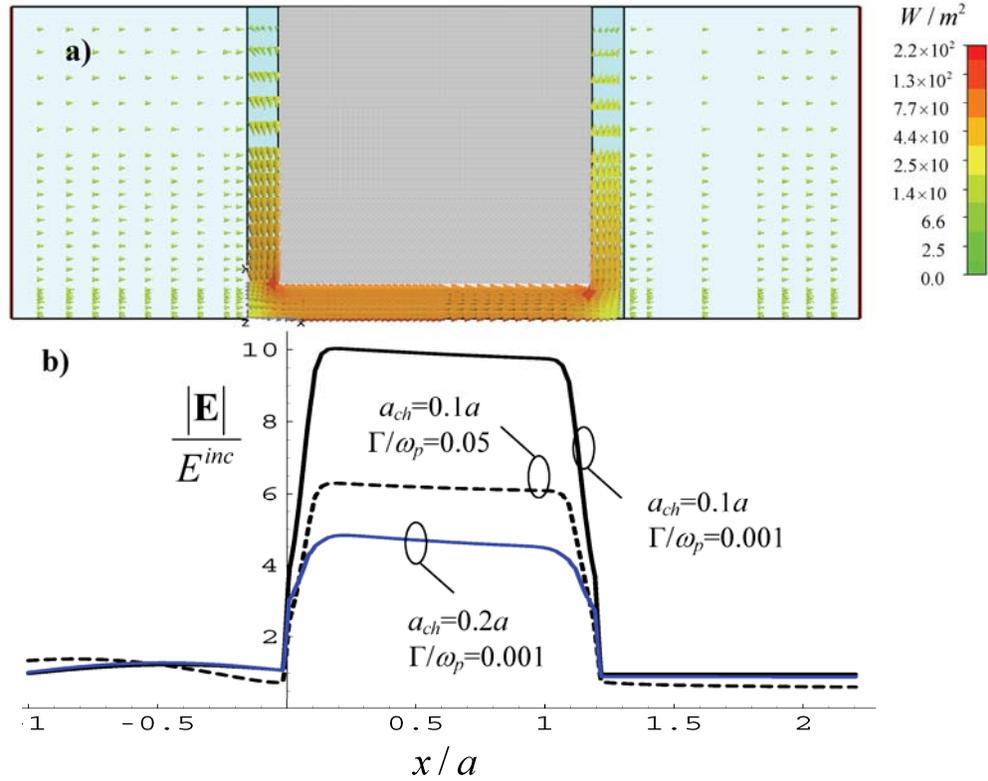

**Fig. 5**. (Color online) Panel (a) Real part of the Poynting vector for $a_{ch} = 0.1a$ and $\Gamma/\omega_p = 0.001$. Panel (b) Electric field (normalized to the amplitude of the incoming wave) along the line $y = 0.5a_{ch}$, for different values of the loss parameter and of the channel height.

The previous results suggest the possibility of using ENZ materials to concentrate the electric field inside a small subwavelength air cavity placed inside the ENZ material. To explore this opportunity, we computed the electromagnetic fields inside the U-shaped channel with an air cavity defined by $0.55a < x < 0.65a$ (see the inset of Fig. 6). The geometry of the channel is as in the previous example and $a_{ch} = 0.1a$. In Fig. 6, we plot the electric and magnetic fields along the line $y = 0.5a_{ch}$. The dashed vertical lines correspond to the interfaces between the air regions and the ENZ material.

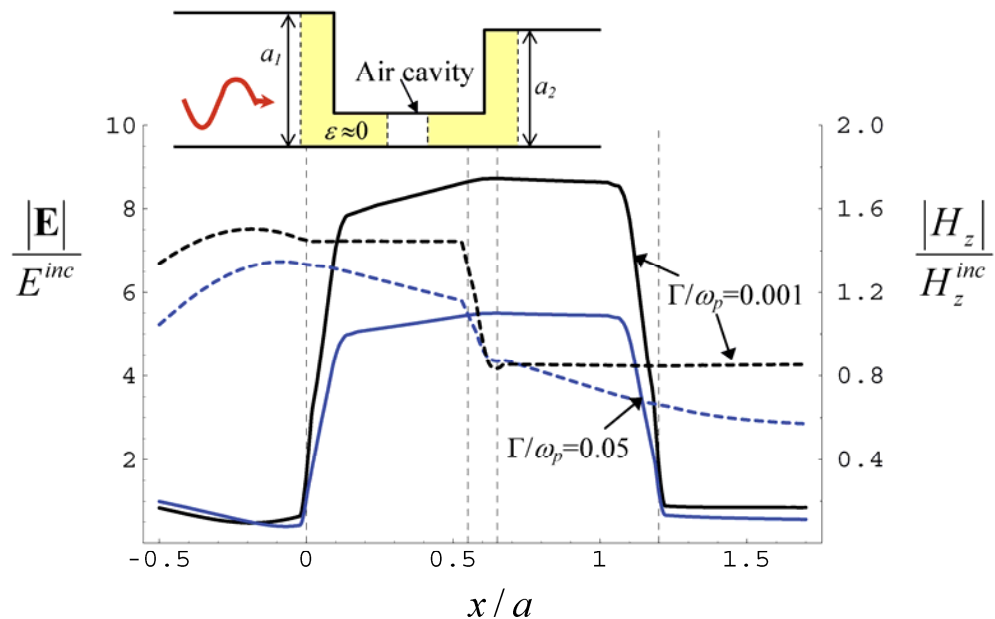

**Fig. 6**. (Color online) Electric field (solid lines) and magnetic field (dashed lines) along the line $y = 0.5a_{ch}$, for different values of the loss parameter. The fields are normalized to the amplitude of the corresponding the incoming wave. The inset represents the geometry of the ENZ channel with an air cavity defined by the region $0.55a < x < 0.65a$, where $a \equiv a_1 = a_2$.

The results of Fig. 6 demonstrate that the electric field may be significantly concentrated inside the air cavity (and also in the ENZ material). It is seen that the electric field inside the cavity is about 9 times (5 times) larger than the incoming field for $\Gamma/\omega_p = 0.001$ ($\Gamma/\omega_p = 0.05$), respectively (these values are only marginally smaller than those reported in Fig. 5 for the case where the transition is completely filled with the ENZ material). It is also worth noting that the magnetic field is nearly constant inside the two sections of the ENZ material for the case $\Gamma/\omega_p = 0.001$, consistently with the theory of [9]. Inside the air cavity the magnetic field may vary appreciably, and to a good approximation is a linear function of *x*. Fig. 5 also shows that when the losses become moderate ($\Gamma/\omega_p = 0.05$), the magnetic field is not anymore uniform inside the ENZ material.

In the $\varepsilon \approx 0$ lossless limit it is possible to calculate the electric field inside the air cavity in closed analytical form. The approach is similar to that of [9] and so the details are omitted here. We only mention that in the $\varepsilon \approx 0$ lossless limit the field inside the air cavity is a superimposition of TEM waves, and that the magnetic field inside each ENZ region is constant. Detailed calculations show that the electric field evaluated at the right-hand side interface of the air cavity with the ENZ material is given by,

$$E_y = E_y^{inc}\Big|_{x=0} 2a_1\left(a_2 - ik_0\mu_{r,p}A_{p2}\right) \Big/ \Big[\cos(k_0 d)a_{ch}\left((a_1+a_2)-ik_0\mu_{r,p}\left(A_{p1}+A_{p2}\right)\right) \\ -i\sin(k_0 d)\left(a_{ch}^2 + a_1 a_2 - ik_0\mu_{r,p}\left(A_{p1}a_2 + A_{p2}a_1\right) - \left(k_0\mu_{r,p}\right)^2 A_{p1}A_{p2}\right)\Big] \quad (3)$$

where $E_y^{inc}$ is the amplitude of the electric field of the incoming wave, *d* is the thickness of the air cavity, and $A_{p1}$ ($A_{p2}$) is the area of the cross-section of the ENZ section at the left (right)–hand side of the air cavity. It can be easily checked that when $a \equiv a_1 = a_2$, $a_{ch}/a \ll 1$, and $k_0\mu_{r,p}A_{p,i}/a \ll 1$ (*i*=1,2), the above formula simplifies to,

$$E_y \approx E_y^{inc}\Big|_{x=0} \frac{1}{\left(\cos(k_0 d)\dfrac{a_{ch}}{a} - i\dfrac{1}{2}\sin(k_0 d)\right)} \quad (4)$$

Thus, provided the cavity is electrically small, $k_0 d \ll 1$, the electric field inside the cavity is enhanced by a factor of $a/a_{ch}$ as compared to the amplitude of the incoming wave, consistently with the results of Fig. 6. These results confirm that ENZ materials may have interesting potentials in concentrating the electric field in a small very subwavelength cavity.

## C. Effect of losses in the metallic walls

In the previous examples it was assumed that the metallic walls were perfect electric conductors. However all metals have losses, and these can be significant at terahertz, infrared and higher frequencies. Thus, it is crucial to evaluate the inevitable effect of metallic losses in the proposed tunneling mechanism. In what follows, we generalize the theory introduced in our previous work [9], so that it can describe the effect of the finite conductivity of the metallic walls.

Here, we consider that the metallic walls are characterized by the bulk conductivity $\sigma$. Thus, the effective permittivity of the metal is of the form:

$$\frac{\varepsilon_{metal}}{\varepsilon_0} \equiv \varepsilon_{r,metal} = 1 - \frac{\sigma}{i\omega\varepsilon_0} \tag{5}$$

We assume that $\sigma/\omega\varepsilon_0 \gg 1$ so that the metals are at least fairly good conductors. We also suppose that $\delta \ll R$, $\delta \ll D$, and $\delta \ll \lambda_{ENZ}/2\pi$, where $\delta = \sqrt{2/\omega\sigma\mu_0}$ is the skin-depth of the metal, $D$ is the thickness of the metallic plates, $R$ is the radius of curvature of the plates at a generic point, and $\lambda_{ENZ}$ is the wavelength in the ENZ material. In these circumstances the interaction of electromagnetic waves with the metallic plates can be described by the Leontovitch impedance boundary condition [16]. The Leontovitch boundary condition assumes that:

$$\mathbf{E}_t = \eta_{metal}\mathbf{J}_c, \qquad \eta_{metal} = \eta_0 \frac{1}{\sqrt{\varepsilon_{r,metal}}} \tag{6}$$

where $\mathbf{J}_c = \hat{\mathbf{v}} \times \mathbf{H}$ the density of current along the metallic walls ($\hat{\mathbf{v}}$ is the unit normal vector directed to the outward of the metallic region), $\mathbf{E}_t$ is the tangential component of the electric field, and $\eta_0$ is the free-space impedance. As proved next, within such an approximation it is possible to calculate in closed analytical form the reflection and transmission coefficients in the $\varepsilon = 0$ lossless limit, and for a generic waveguide scenario.

To begin with, consider a generic waveguide configuration as in [9] where an ENZ filled channel is illuminated by the fundamental TEM waveguide mode (the geometry of Fig. 1 corresponds to the particular case in which the ENZ transition has the U-shape). As proved in [9], in the $\varepsilon = 0$ lossless limit the magnetic field is necessarily constant in the ENZ material, $H_z = H_z^{\text{int}}$, and thus, as discussed in the end of section II.A, the density of current along the metallic walls also has constant amplitude: $\mathbf{J}_c = H_z^{\text{int}} \hat{\mathbf{t}}$ where $\hat{\mathbf{t}} = \hat{\mathbf{v}} \times \hat{\mathbf{u}}_z$ is the unit vector tangent to the metallic surface. As in [9], next we apply Faraday's law to the counterclockwise contour that encloses the ENZ region. We obtain that

$$\int_{\substack{\text{interfaces} \\ \text{with air}}} \mathbf{E}.\mathbf{dl} + \int_{\text{metal walls}} \mathbf{E}.\mathbf{dl} = +i\omega\mu_0\mu_{r,p}H_z^{\text{int}} A_p, \qquad (7)$$

where $A_p$ is the cross-sectional area of the ENZ channel, and the first line integral is over the interfaces with air ($x = 0$ and $x' = 0$ in Fig. 1) whereas the second line integral is over the metallic walls. Using the Leontovitch boundary condition (6) and $\mathbf{J}_c = H_z^{\text{int}} \hat{\mathbf{t}}$, it is clear that $\int_{\text{metal walls}} \mathbf{E}.\mathbf{dl} = \eta_{metal} H_z^{\text{int}} L_{tot}$ where $L_{tot}$ is the total length of the metallic plates enclosing the ENZ material (i.e. the sum of the lengths of the two plates; for the particular geometry of Fig. 1, we have $L_{tot} = 2(L_1 + L + L_2) + (a_1 + a_2 - 2a_{ch})$). On the other hand, following [9], it is simple to verify that

$$\int_{\substack{\text{interfaces} \\ \text{with air}}} \mathbf{E} \cdot \mathbf{dl} = \eta_0 H_z^{\text{int}} a_2 - \eta_0 H_z^{\text{int}} \frac{1-\rho}{1+\rho} a_1,$$ where $\rho$ is the reflection coefficient for the magnetic field. Substituting these results in (7) we readily obtain that:

$$\rho = \frac{(a_1 - a_2) - \left(-i k_0 \mu_{r,p} A_p + \frac{1}{\sqrt{\varepsilon_{r,metal}}} L_{tot}\right)}{(a_1 + a_2) + \left(-i k_0 \mu_{r,p} A_p + \frac{1}{\sqrt{\varepsilon_{r,metal}}} L_{tot}\right)} \quad (8)$$

The transmission coefficient is given by $T = 1 + \rho$. The above formula is the generalization of (1) to the case where the metallic walls have finite conductivity. We underline that the formula is completely general (valid for channels with nearly arbitrary geometry), and only assumes the conditions implicit in the Leontovitch boundary condition approximation and that $\varepsilon = 0$ in the ENZ material. Equation (8) predicts that for the case $a \equiv a_1 = a_2$ a wave may tunnel through the ENZ channel only if $k_0 \mu_{r,p} A_p / a \ll 1$ (which was the condition derived in our previous work [9]) and if $L_{tot} / \left(a \sqrt{|\varepsilon_{r,metal}|}\right) \ll 1$. This latter condition is very interesting because it establishes that the effect of losses depends mostly on the ratio $L_{tot}/a$, but not specifically on the distance between the two metallic plates. This result implies that compressing the transverse section of the ENZ channel more and more, does not in principle affect the loss in the metallic walls. This somewhat unexpected property can be understood by noting that the magnetic field (and consequently $\mathbf{J}_c$) is not enhanced inside the ENZ channel as compared to the amplitude of the incident wave (indeed, only the electric field is greatly enhanced in the channel, as illustrated in Fig. 5). Thus, since the losses in the metal are proportional to $\mathbf{J}_c$, it is clear that squeezing the transverse cross-section of the ENZ channel (with respect to the direction of propagation) does not necessarily result in an increase of metallic losses. Moreover, the condition $L_{tot}/\left(a\sqrt{|\varepsilon_{r,metal}|}\right) \ll 1$ also demonstrates that for values of $|\varepsilon_{r,metal}|$ moderately large the effect of metallic losses may be negligible. For example, if $L_{tot}/a = 5$ the effect of losses in the metallic walls may be negligible for $|\varepsilon_{r,metal}| > 100$. To give an idea of the possibilities we suggest here, we note that for example aluminum (Al) may satisfy this condition for $\lambda_0 > 850nm$ [17].

In order to confirm the proposed theory for the loss in the walls, we used CST Microwave Studio$^{TM}$ to compute the S-parameters of a U-shaped channel (Fig. 1) with walls with finite conductivity $\sigma$. The geometry of the channel is as in the first example of section II.A (in particular, $L = 1.0a$, $a_{ch} = L_1 = L_2 = 0.1a$ and $\omega_p a/c = \pi/2$). To fully assess the effect of loss in the walls, we considered $\Gamma/\omega_p = 0.001$ in the ENZ material (i.e. the dielectric losses are negligible). In Fig. 7 the transmission coefficient is depicted as a function of frequency for different values of $\sigma$. The star symbols represent the results yielded by Eq. (8) (the transmission coefficient is $T = 1 + \rho$ at $\omega = \omega_p$). It is verified that the star symbols agree well with the full wave results, even for values of $\sigma$ moderately small. This validates our theory and the use of the Leontovitch impedance boundary condition. The results also illustrate that the transmission coefficient is not very much affected by metallic losses, even for $\dfrac{\sigma}{\omega_p \varepsilon_0} = 13.9$, which corresponds to $|\varepsilon_{r,metal}| \approx 13.9$. Hence, we conclude that in practice the effect of losses in the metallic walls may be of second order, and that in principle the loss in the ENZ material (discussed in section II.A) is the most relevant source of loss in the system under study.

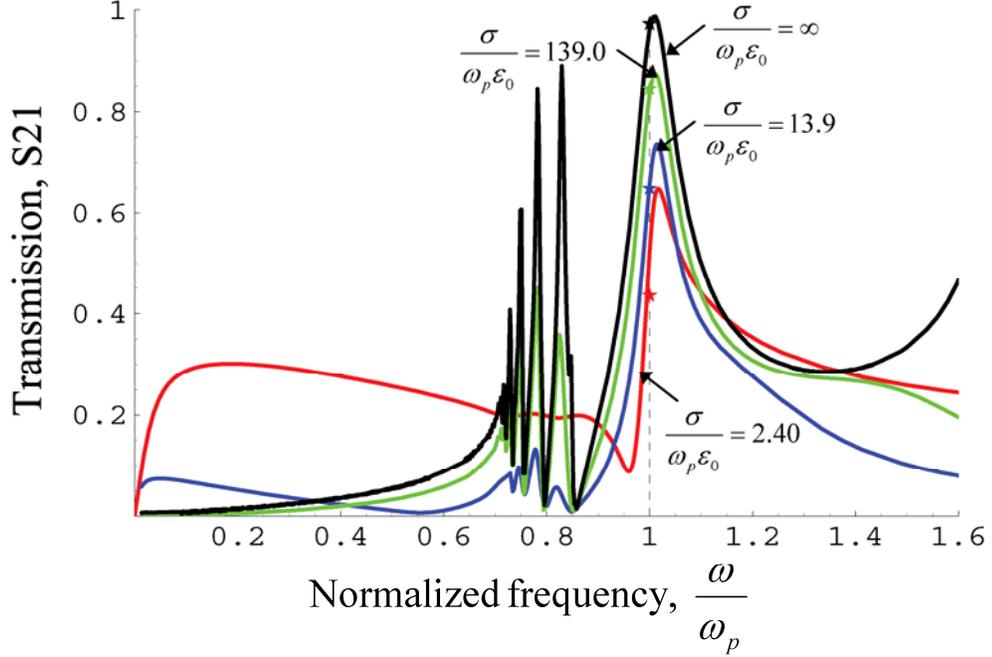

**Fig. 7**. (Color online) Amplitude of the transmission coefficient (S21-parameter) as a function of normalized frequency for a U-shaped ENZ channel and different values of the conductivity of the walls. The "star" symbols along the vertical line $\omega = \omega_p$ represent the results computed using formula (8).

## D. Group velocity

It is interesting to analyze the group velocity for a packet of electromagnetic waves that tunnels through an ENZ filled channel. We consider for simplicity that the dielectric permittivity $\varepsilon$ of the ENZ material is characterized by a Drude-type dispersion model, $\varepsilon = 1 - \omega_p^2 / \omega(\omega + i\Gamma)$. It is well known, that in the lossless case ($\Gamma = 0$) the phase velocity is $v_p = c/\sqrt{\varepsilon}$, whereas the group velocity in the unbounded medium is $v_g = \sqrt{\varepsilon} c$ ($c$ is the speed of light in vacuum). Thus, in the lossless $\varepsilon = 0$ limit the phase velocity is $v_p = \infty$ whereas the group velocity is $v_g = 0$. This result seems to contradict the possibility of transmitting a signal through an ENZ filled channel. However, a more careful analysis shows that in fact that is not the case. The key point is that $v_g = \sqrt{\varepsilon} c$ represents the group velocity of a packet of waves that propagates in an *unbounded* ENZ material, and thus this formula does not necessarily apply for propagation in a finite thickness ENZ sample, where significant coupling between the two interfaces may effectively permit that a packet of waves is squeezed through the ENZ channel.

In fact, consider again the geometry of Fig. 1 where the TEM fundamental waveguide mode illuminates a U-shaped ENZ channel. In general, the transmitted wave (after it emerges from the ENZ transition) may consist of a superposition of waveguide modes. However, as proved in [9], when $\varepsilon = 0$ no evanescent modes are excited in the region $x' > 0$, and thus it is a good approximation to consider that in the ENZ limit the transmitted wave consists only of a TEM mode. Let $T = T(\omega)$ be the corresponding transmission coefficient so that the amplitude of the transmitted magnetic field at $x' = 0$, $H_z^{tx}(\omega)$, is related with the incident field at $x = 0$, $H_z^{inc}(\omega)$, through the relation $H_z^{tx}(\omega) = T(\omega) H_z^{inc}(\omega)$. Let $A(\omega)$ and $\phi(\omega)$ be the amplitude and phase of the transmission coefficient so that $T(\omega) = A(\omega) e^{i\phi(\omega)}$. Using a Taylor expansion for the phase around $\omega = \omega_0$ we can write that:

$$T(\omega) \approx A(\omega_0) e^{i\theta_0} e^{i \frac{d\phi}{d\omega} \omega}, \qquad \theta_0 = \phi(\omega_0) - \left.\frac{d\phi}{d\omega}\right|_{\omega=\omega_o} \omega_0 \qquad (9)$$

Suppose now that the incoming plane wave is of the form $\tilde{H}_z^{inc}(t) = H(t) \cos(\omega_0 t)$ where $H(t)$ is a slow-varying function of time (i.e. the Fourier transform of $\tilde{H}_z^{inc}(t)$ – $H_z^{inc}(\omega)$ – corresponds to a packet waves with spectrum concentrated around $\omega = \omega_0$). A straightforward analysis shows that within the approximation (9), the magnetic field at the output plane $x' = 0$ is given by:

$$\tilde{H}^{tx}(t) \approx A(\omega_0) H(t-t_d) \cos(\omega_0(t-t_d) - \theta_0), \qquad t_d = \left.\frac{d\phi}{d\omega}\right|_{\omega=\omega_o} \tag{10}$$

Thus, the envelope of the incoming field is reproduced at the output plane with a delay of $t_d$ unities of time (i.e. $t_d$ is the time necessary for the packet of waves to travel across the ENZ channel). We define the group velocity as $v_g = \frac{d}{t_d}$, where $d$ is the length of the ENZ channel (e.g. for the geometry of Fig. 1, $d = L_1 + L + L_2$). Hence, we find that $v_g = d \left(\frac{d\phi}{d\omega}\right)^{-1}\bigg|_{\omega=\omega_o}$, which can be rewritten as:

$$\frac{v_g}{c} = d \left(\frac{d\phi}{dk_0}\right)^{-1}\bigg|_{\omega=\omega_o} \tag{11}$$

where $k_0 = \omega\sqrt{\varepsilon_0\mu_0}$ is the free-space wave number. We will use the above formula to calculate $v_g$ inside the ENZ channel. Again, we underline that (11) is not equivalent to the result $v_g = \sqrt{\varepsilon} c$, which represents the group velocity inside an unbounded medium, and thus, in general, does not characterize $v_g$ in the ENZ channel.

In order to estimate the group velocity for a U-shaped channel (Fig. 1), we computed the transmission coefficient $T(\omega)$ using a full wave electromagnetic simulator [14], and then we evaluated $v_g$ using (11). In the simulations, it was assumed that $L_1 = L_2 = 0.1a$ and $\omega_p a/c = \pi/2$, consistently with the parameters chosen in section II.A. The results for $a_{ch} = 0.1a$ and $L = 1.0a$ are depicted in Fig. 8 for $\Gamma/\omega_p = 0.001$ (curve *a*) and $\Gamma/\omega_p = 0.05$ (curve *b*). It is seen that the group velocity has a minimum around the plasma frequency $\omega = \omega_p$. For $\Gamma/\omega_p = 0.001$ the group velocity can be as small as $v_g = 0.08c$. It is also verified that the group velocity is improved (around $\omega = \omega_p$) when the losses are increased to $\Gamma/\omega_p = 0.05$ (curve *b*), and also for $\omega < \omega_p$. However, we should point out that the group velocity may lose its meaning if the amplitude of the transmission coefficient becomes close to zero [18, pp. 325], and thus the results for $\omega$ moderately below $\omega_p$ may not be applicable (see Fig. 2).

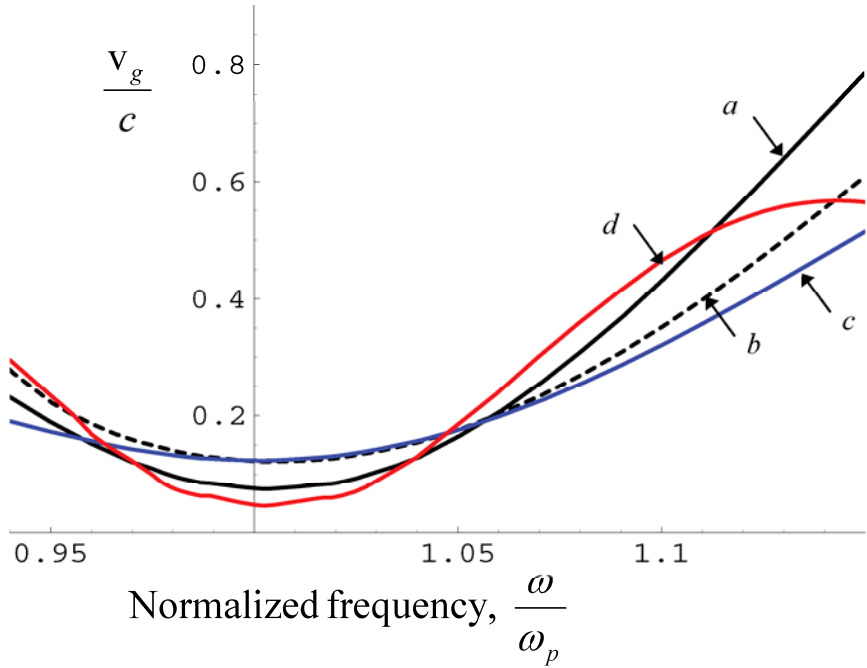

**Fig. 8**. (Color online) Group velocity as a function of the normalized frequency. The collision (damping) frequency is such that $\Gamma/\omega_p = 0.001$, except for curve (*b*) where $\Gamma/\omega_p = 0.05$. Curves (*a*) (solid black line) and (*b*) (dashed black line): $a_{ch} = 0.1a$ and $L = 1.0a$. Curve (*c*) (blue line – dark gray in grayscale): $a_{ch} = 0.2a$ and $L = 1.0a$. Curve (*d*) (red line – light gray in grayscale): $a_{ch} = 0.1a$ and $L = 2.0a$.

The results of Fig. 8 also demonstrate that increasing the channel height $a_{ch}$ (curve *c*) results in an increase of the group velocity around $\omega = \omega_p$, even though the transmissivity of the channel is slightly worsened. On the other hand, increasing the channel length *L* (curve *d*) results in a decrease of the group velocity as one could intuitively expect.

## *E. Plasma simulation at microwaves*

The simulations and results presented in the previous subsections consider that the problem under study is intrinsically two-dimensional (2D), and that the ENZ material is readily available in nature. In what follows, we prove that it is possible to emulate such propagation scenarios at microwaves, and reproduce the supercoupling, field concentration, and tunneling effects using a realistic three-dimensional (3D) setup that only requires using standard dielectrics and metals with an empty section!

The main idea is to simulate the 2D artificial plasma using two parallel metallic plates co-planar with the electric field and with the plane of propagation (i.e. parallel to the *x-y* plane). It is known [19, 20, 21] that such configuration may mimic the behavior of a plasmonic material with the Drude-type dispersion as $\varepsilon = \varepsilon_{d,r} - (\pi/k_0 s)^2$, where $\varepsilon_{d,r}$ is the relative permittivity of the dielectric between the plates, and *s* is the distance between the plates. Such concepts were used in our previous work [10] to emulate a medium with $\varepsilon$ and $\mu$ simultaneously near zero at microwaves. Here, we use similar concepts to demonstrate the supercoupling, field confinement, and tunneling effect under study here at microwaves.

In the first example, we analyze a closed 3D metallic waveguide environment. We consider that the cross-section of the waveguide (contained in the *x-y* plane) is as shown in Fig. 1, i.e. the metallic walls have the U-shape. In order to simulate an artificial plasma two parallel-plates are inserted at the planes $z=0$ and $z=s$. As discussed in [10], the 3D metallic waveguide is filled with two different dielectrics, so that the artificial plasma may emulate (at the design frequency) a dielectric with $\varepsilon_{eff,air}=1.0$ in the "free-space regions" ($x<0$ or $x'>0$ in Fig. 1), and an ENZ material with $\varepsilon_{eff,ENZ}=0$ in the ENZ transition. Let us first discuss the design of the ENZ transition. As in the previous sections, we choose the plasma frequency such that $\omega_p a/c = \pi/2$. Thus, supposing that the region that mimics the behavior of the ENZ transition is filled with air, i.e., an empty region with ($\varepsilon_{d,r}=1$), from $\varepsilon_{eff,ENZ}=0=\varepsilon_{d,r}-\left(\pi/k_0 s\right)^2$ we find that the required distance between the plates (along the z-direction) is $s=2a$. Let us discuss now, the design of the waveguide regions that are supposed to behave as a medium with $\varepsilon_{eff,air}=1.0$. At the normalized frequency $\omega_p a/c=\pi/2$ and for $s=2a$, the effective permittivity of the simulated plasma is $\varepsilon_{eff,air}=\varepsilon_{d,r}-1$. Thus, in order that $\varepsilon_{eff,air}=1.0$ (at the design frequency), we must choose $\varepsilon_{d,r}=2.0$ in the regions $x<0$ or $x'>0$ of Fig. 1. In conclusion, in order to emulate the behavior of the 2D waveguide configuration depicted in Fig. 1, we consider a 3D closed metallic waveguide with the same cross-section in the x-y plane and with metallic walls at $z=0$ and $z=2a$. The region that is supposed to behave as the ENZ transition is filled with air ($\varepsilon_{d,r}=1.0$), whereas the regions that are supposed to mimic the behavior of the free-space sections in Fig. 1 are filled with a dielectric with $\varepsilon_{d,r}=2.0$. The 3D waveguide is excited with the fundamental TE10 mode with electric field in the *xoy*-plane. This mode effectively emulates the behavior of the TEM mode in 2D-configuration of Fig. 1 around $\omega \approx \omega_p$. Note that the TE10 mode is cut-off for $\omega < 0.7\omega_p$ (for $\omega = 0.7\omega_p$ the 3D waveguide sections filled with $\varepsilon_{d,r} = 2.0$ emulate an ENZ material, while the unfilled region with $\varepsilon_{d,r} = 1.0$ emulates a material with negative permittivity).

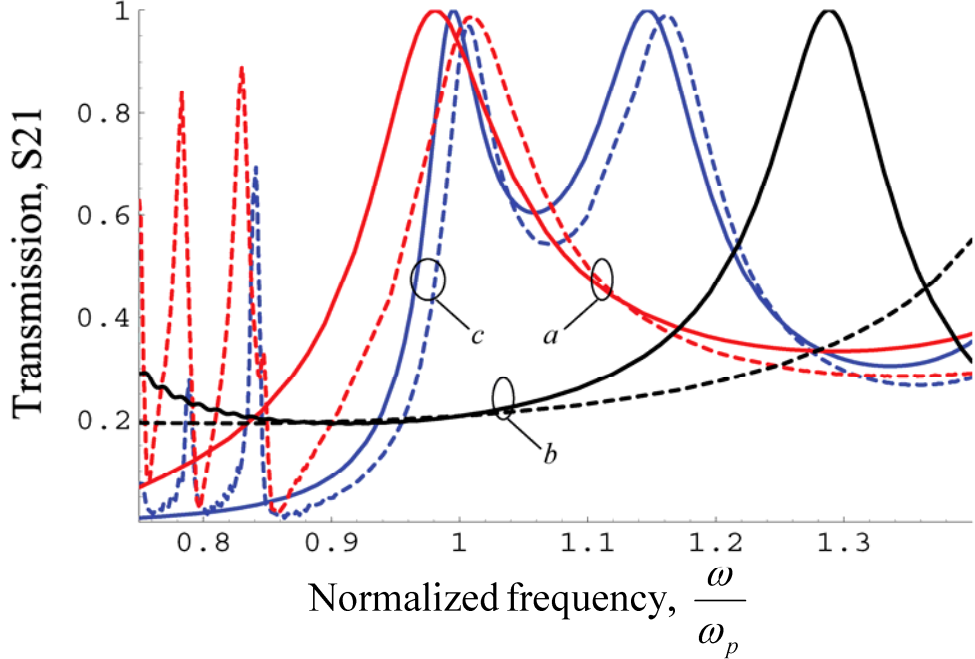

**Fig. 9**. (Color online) S21-parameter as a function of normalized frequency for an artificial plasma (solid lines) emulated using a 3D-metallic waveguide, and for the corresponding 2D-setup of Fig. 1 (dashed lines). (a) U-shaped transition with $L = 1.0a$. (b) 3D-waveguide uniformly filled with $\varepsilon_{d,r} = 2.0$ (solid line) and the corresponding 2D-setup with the narrow channel filled with air (dashed line). (c) Similar to (a) but with $L = 3.0a$.

In Fig. 9 we depict the calculated S21 parameter as a function of frequency for several waveguide configurations with $L_1 = L_2 = 0.1a$ and $a_{ch} = 0.1a$. The solid lines associated with curves (a) and (c) correspond to 3D setups with artificial plasmas with $L = 1.0a$ and $L = 3.0a$. The corresponding dashed lines are the results obtained for the equivalent 2D-geometry, where the narrow channel is filled with an ideal low loss ENZ material. It is seen that around $\omega = \omega_p$, the transmission characteristic associated with the artificial plasma (emulated with a 3D metallic waveguide filled with standard dielectrics and empty region) mimics closely (apart from a slight shift in frequency) the transmission characteristic of the corresponding 2D-ENZ filled narrow channel. In particular, it is confirmed that around $\omega = \omega_p$ the wave tunnels through the narrow channel, independent of its specific length. For comparison, we also plot in curve (b) the transmission characteristic of a 3D waveguide with the same geometry as in curve (a), but uniformly filled with a dielectric with $\varepsilon_{d,r} = 2.0$. This setup is supposed to emulate the behavior of the associated 2D-waveguide (Fig. 1) when the narrow channel filled with air. The results of Fig. 9 indeed demonstrate that the transmission characteristic of both configurations is very similar at $\omega \approx \omega_p$. As mentioned before, for frequencies significantly different from $\omega_p$ the simulated plasma is not expected to emulate the configuration of Fig. 1 so well, because the effective permittivity of the regions filled with $\varepsilon_{d,r} = 2.0$ is not anymore $\varepsilon_{eff,air} = 1.0$ due to the dispersive behavior of the equivalent material.

In a second example, we demonstrate a similar supercoupling and tunneling effect in a 3D microstrip line configuration. The geometry of the problem is shown in panel (a) of Fig. 10. The microstrip line consists of a conducting ground plane, a dielectric slab with thickness $a$, and conducting strip with width $W$. It is well known that for low frequencies such open waveguide structure supports a quasi-TEM wave, with a modal structure similar to the TEM mode supported by a parallel-plate 2D waveguide. We consider that the permittivity of the dielectric slab is $\varepsilon_s = 2.2$ and $W = 4a$ so that the line impedance of the microstrip line is $Z_c = 43\,[Ohm]$ (calculated using CST Microwave Studio$^{TM}$ [14]).

It is supposed that the microstrip conductor has an abrupt transition shaped as an "U" (Fig. 10). The transverse cross-section of the line (relative to the $z$-direction) has geometry similar to that of Fig. 1. As in previous simulations, we choose $L_1 = L_2 = a_{ch} = 0.1a$ and $L = 1.0a$.

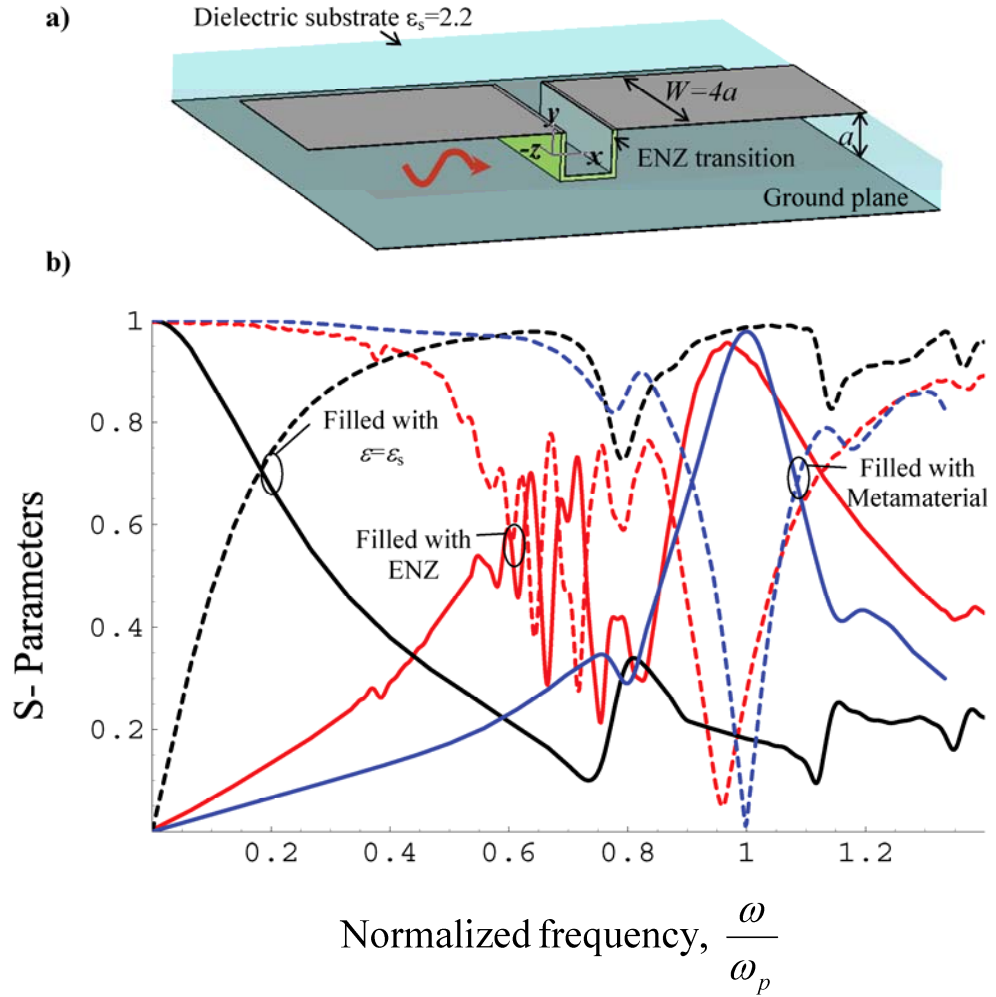

**Fig. 10**. (Color online) Panel (a) Microstrip line with a U-shaped transition filled with an ENZ metamaterial. The microstrip line is illuminated with the quasi-TEM transmission line mode with electric field along $y$ and magnetic field along $z$. The cut of the structure along the center of the microstrip conductor has geometry similar to that of Fig. 1. Panel (b) S11-parameter (dashed lines) and S21-parameter (solid lines) as a function of normalized frequency for a transition filled with (i) dielectric with $\varepsilon = \varepsilon_s$

(black lines), (ii) low-loss ENZ material (blue lines - dark gray in grayscale), (iii) artificial plasma that emulates ENZ material (red lines – light gray in grayscale).

From the results of previous sections, it is expected that if the U-shaped transition is filled with the same material as the dielectric substrate ($\varepsilon_s = 2.2$), most of the energy is reflected at the U-transition for moderately large frequencies. This is confirmed in panel (b) of Fig. 10, where we plot the calculated S11- (reflection) and S21- (transmission) parameters as a function of frequency (black dashed and solid lines, respectively). In order to improve the transmissivity of the microstrip configuration, we may fill the U-shaped transition with an ENZ material, as illustrated in panel (a) of Fig. 10. In the simulations we chose the plasma frequency such that $\omega_p a \sqrt{\varepsilon_s} / c = 1.16$ (this value was chosen to simplify the design of the artificial plasma, as described ahead), and the damping frequency $\Gamma = 0.01 \omega_p$. The corresponding S-parameters calculated using the full wave simulator [14] are depicted in Fig. 10 (red lines). It is seen that consistently with the results obtained for the geometry of Fig. 1, the wave may tunnel through the narrow channel around the plasma frequency. As described next, an artificial plasma realization of the ENZ material may also enable a similar tunneling phenomenon. As in the first example of this subsection, the simulated plasma is designed using a parallel-plate configuration. In this way, the microstrip conductor is short-circuited to the ground plane through two parallel metallic plates (normal to the z-direction) placed at the U-shaped transition. The metallic plates are located at the planes $z = -W/2$ and $z = +W/2$, supposing that the conducting strip is defined by the region $-W/2 < z < W/2$. The length of the metallic plates (along the x-direction) is $L_1 + L + L_2$ (see Fig. 1). The space in between the two metallic plates (U-shaped transition) is filled with a dielectric with permittivity $\varepsilon_{d,r}$. It is expected that this simple structure emulates an artificial plasma characterized by $\varepsilon_{eff,ENZ} = \varepsilon_{d,r} - (\pi/k_0 W)^2$. Choosing for example $\varepsilon_{d,r} = 1.0$, we find that $\varepsilon_{eff,ENZ} = 0$ for $\omega_p$ such that $\omega_p a \sqrt{\varepsilon_s}/c = 1.16$ (this justifies our choice for the plasma frequency). The corresponding S-parameters calculated for a microstrip line with an artificial plasma (metamaterial) transition are depicted in Fig. 10 (blue lines). The results demonstrate that consistently with our theory the wave may completely tunnel through the U-shaped artificial plasma transition. It is worth noting that the results for the artificial plasma realization are even slightly better than those obtained for the ideal ENZ material, perhaps because the effect of losses in the simulated plasma is negligible. These results confirm that the described supercoupling tunneling effect may, in fact, be demonstrated at microwave frequencies using a realistic 3D setup, which only involves standard dielectric and conducting materials with arguably lower material losses.

## III. Supercoupling and Squeezing Energy through ENZ Anisotropic Channels

As discussed in the Introduction and in [9], even though ENZ materials may readily available in nature at certain specific frequencies of the electromagnetic spectrum, in the general case we may need to synthesize them as microstructured materials (metamaterials). Since anisotropic materials may be easier to fabricate as compared to isotropic materials, in this section we investigate the potentials of anisotropic materials with permittivity near zero. In addition, we discuss the realization of these artificial materials using wire media.

## A. Characterization of ENZ Anisotropic Channels

In this subsection, we extend the theory developed in [9] to the case in which the ENZ material is anisotropic. We will prove that when the metallic channel is filled with an anisotropic material such that the permittivity component normal to the interface is near zero, then, in certain conditions, the scattering problem may be solved in closed analytical form. Moreover, we will show that for certain geometries the scattering parameters may be made independent of the specific geometry of the channel, in analogy with the results obtained in [9].

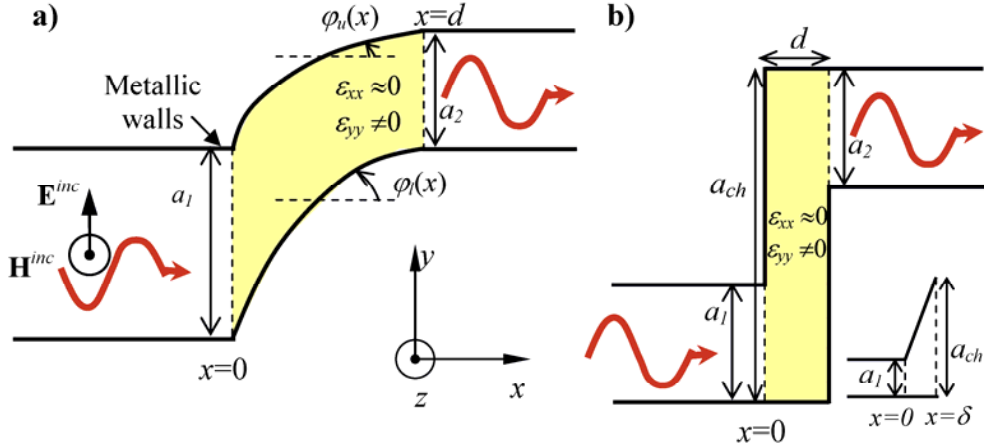

**Fig. 11**. (Color online) Panel (a): Geometry of a metallic waveguide with an abrupt transition filled with an anisotropic ENZ material. The incident wave is the fundamental TEM mode. $\varphi_u(x)$ and $\varphi_l(x)$ define the angles between the vector tangent to the upper/lower metallic plates and the $x$-direction. Panel (b): similar to panel (a) for the case of a transition with two 90[deg]-bends and thickness $d$. The small inset in panel (b) represents the detail of the interface $x=0$ where the profile of the upper plate is varied continuously over an infinitesimal length $\delta$ (see the text).

The geometry of the problem is depicted in panel (a) of Fig. 11. Two parallel-plate metallic waveguides are interfaced by an abrupt transition filled with an ENZ anisotropic material. We suppose that the structure is invariant to translations along the z-direction and that $\partial/\partial z = 0$, so that the problem is effectively two-dimensional. In the region $0 < x < d$ the profile of the upper metallic plate is defined parametrically by $(x, y_u(x))$, whereas the profile of the lower plate is defined by $(x, y_l(x))$. The angles between the associated tangent vectors and the x-direction are $\varphi_u(x)$ and $\varphi_l(x)$, as illustrated in Fig. 11. The incident wave (propagating in the region $x < 0$) is the fundamental TEM mode. The ENZ material is characterized by the (relative) permittivity dyadic:

$$\bar{\bar{\varepsilon}} = \varepsilon_{xx}\hat{\mathbf{u}}_x\hat{\mathbf{u}}_x + \varepsilon_{yy}\hat{\mathbf{u}}_y\hat{\mathbf{u}}_y + \varepsilon_{zz}\hat{\mathbf{u}}_z\hat{\mathbf{u}}_z \tag{12}$$

The objective is to characterize the reflection and transmission parameters in the regime where $\varepsilon_{xx} \approx 0$, with $\varepsilon_{yy}$ not necessarily near zero (the $\varepsilon_{zz}$ component is not relevant here since for the geometry under study the electric field is confined to the x-y plane). The electromagnetic fields are assumed H-polarized with $\mathbf{H} = H_z \hat{\mathbf{u}}_z$. The corresponding electric field is given by:

$$\mathbf{E} = \frac{1}{-i\omega\varepsilon_0}\left(\frac{1}{\varepsilon_{xx}}\frac{\partial H_z}{\partial y}, -\frac{1}{\varepsilon_{yy}}\frac{\partial H_z}{\partial x}, 0\right) \tag{13}$$

It is straightforward to verify that $H_z$ satisfies the equation:

$$\frac{\partial}{\partial x}\frac{1}{\varepsilon_{yy}}\frac{\partial H_z}{\partial x} + \frac{\partial}{\partial y}\frac{1}{\varepsilon_{xx}}\frac{\partial H_z}{\partial y} + k_0^2 \mu_{r,p} H_z = 0 \tag{14}$$

where $\mu_{r,p}$ is the relative permeability (along the z-direction) of the ENZ material, and $k_0 = \omega\sqrt{\varepsilon_0\mu_0}$. Next we prove that in the $\varepsilon_{xx} = 0$ limit the magnetic field inside the ENZ material is such that:

$$H_z = H_z(x), \quad \text{in the } \varepsilon_{xx} = 0 \text{ anisotropic material} \tag{15}$$

i.e. the magnetic field is independent of the y-coordinate. In fact, from the Poynting's theorem [18, pp. 265] we may easily obtain that:

$$i\omega\varepsilon_0 \int_{\partial A_{ENZ}} (\hat{\mathbf{v}} \times \mathbf{E}) \cdot \hat{\mathbf{u}}_z H_z^* dl = \int_{A_{ENZ}} \frac{1}{\varepsilon_{yy}} \left| \frac{\partial H_z}{\partial x} \right|^2 + \frac{1}{\varepsilon_{xx}} \left| \frac{\partial H_z}{\partial y} \right|^2 - k_0^2 \mu_{r,p} |H_z|^2 d^2\mathbf{r} \qquad (16)$$

where $A_{ENZ}$ represents the ENZ region, $\partial A_{ENZ}$ is the corresponding boundary (a closed contour), and $\hat{\mathbf{v}}$ is the outward unit vector. Next, we note that since the tangential electromagnetic fields are continuous across the ENZ material-air interface, then the left-hand side integral is expected to be uniformly bounded in the $\varepsilon_{xx} = 0$ limit (i.e. the electromagnetic fields in the air region are expected to remain uniformly bounded, independent of $\varepsilon_{xx}$). On the other hand, supposing that $\mu_{r,p}$ and $\varepsilon_{yy}$ are different from zero and have a positive imaginary component that takes into account small losses (which may be negligibly small), it follows that (calculating the imaginary part of both members of (16)) the integrals corresponding to the first and third terms in the integrand of the right-hand side member are uniformly bounded in the $\varepsilon_{xx} = 0$ limit. Thus, we conclude that both the left-hand side member and the integrals corresponding to the first and third terms in the integrand of the right-hand side member are uniformly bounded in the $\varepsilon_{xx} = 0$ limit. Hence, multiplying both sides of (16) by $\varepsilon_{xx}$ and letting $\varepsilon_{xx} \to 0$ it follows that $\int_{A_{ENZ}} \left| \frac{\partial H_z}{\partial y} \right|^2 d^2\mathbf{r} \to 0$, and consequently $\frac{\partial H_z}{\partial y}$ vanishes inside the ENZ material. This implies that (15) holds, as we wanted to prove. It is also worth pointing out that condition (15) is necessary in order that the electric field inside the ENZ material may be finite in the $\varepsilon_{xx} = 0$ limit. Note that if the material was isotropic and $\varepsilon_{xx} = \varepsilon_{yy} = 0$, then using a similar reasoning we would obtain $\nabla H_z = 0$, which would imply $H_z = const.$, consistently with the results of [9].

Note that (15) implies that the ENZ anisotropic channel "freezes" the field variations along the *y*-direction, and thus enforces that the wavefronts inside the ENZ channel are parallel to the interfaces with air. In particular, for TEM incidence with $H_z^{inc} = H_0^{inc} e^{+ik_0 x}$, it follows that the reflected and transmitted waves must also be TEM (i.e. in the $\varepsilon_{xx} = 0$ limit it is not possible to excite evanescent TM$^x$ modes inside the air regions). Thus, we can write that:

$$H_z = H_0^{inc}\left(e^{ik_0 x} + \rho e^{-ik_0 x}\right), \qquad x < 0 \qquad (17a)$$

$$H_z = H_0^{inc} T e^{+ik_0(x-d)}, \qquad x > d \qquad (17b)$$

where $\rho$ and $T$ are the reflection and transmission coefficients for the magnetic field, respectively. In order to calculate the scattering parameters we need to determine $H_z$ inside the ENZ material. To this end, we integrate both sides of Eq. (14) over $y \in [y_l, y_u]$. Because of (15) both $H_z$ and its derivatives in *x* are independent of *y* in the $\varepsilon_{xx} = 0$ limit. Thus, we find that:

$$\frac{\partial}{\partial x}\frac{1}{\varepsilon_{yy}}\frac{\partial H_z}{\partial x} + k_0^2 \mu_{r,p} H_z = -\frac{1}{y_u - y_l}\frac{1}{\varepsilon_{xx}}\frac{\partial H_z}{\partial y}\bigg|_{y_l}^{y_u} \qquad (18)$$

Note that the term $\dfrac{1}{\varepsilon_{xx}}\dfrac{\partial H_z}{\partial y}$ is indeterminate in the $\varepsilon_{xx} = 0$ limit. However, noting that the tangential electric field must vanish at the metallic walls, it is clear that for arbitrary $\varepsilon_{xx}$ we must have $\dfrac{1}{\varepsilon_{xx}}\dfrac{\partial H_z}{\partial y}\bigg|_{y=y_i(x)} = \dfrac{1}{\varepsilon_{yy}}\dfrac{\partial H_z}{\partial x}\bigg|_{y=y_i(x)} \tan\varphi_i$, *i=u,l* , (see panel (a) of Fig. 11).

Thus, we conclude that in the $\varepsilon_{xx} = 0$ limit $H_z = H_z(x)$ satisfies the following ordinary differential equation:

$$\frac{\partial}{\partial x}\frac{1}{\varepsilon_{yy}}\frac{\partial H_z}{\partial x} + \frac{\tan\varphi_u - \tan\varphi_l}{y_u - y_l}\frac{1}{\varepsilon_{yy}}\frac{\partial H_z}{\partial x} + k_0^2 \mu_{r,p} H_z = 0, \qquad \text{for } \varepsilon_{xx} = 0 \qquad (19)$$

By solving the differential equation (which in general depends on the specific geometry of the problem) we can obtain $H_z = H_z(x)$ inside the ENZ anisotropic region, and afterwards using (17) the unknown scattering parameters $\rho$ and $T$.

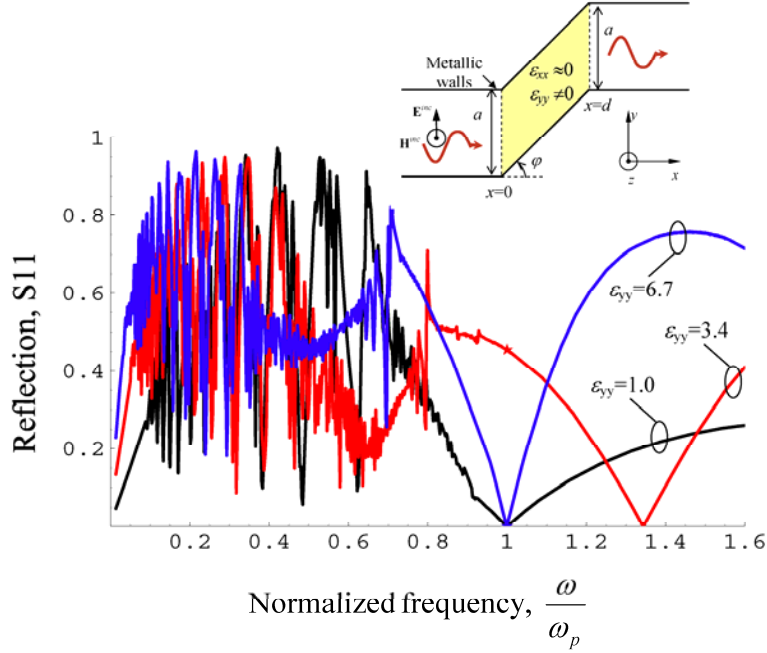

**Fig. 12.** (Color online) Reflection characteristic of a waveguide filled with an ENZ anisotropic material as a function of normalized frequency. The star symbols at $\omega = \omega_p$ represent the theoretical values computed using (21). The distance between the upper and lower plates is uniform in the ENZ section. The inset represents the geometry of the structure.

In order to illustrate this procedure and the application of the formalism let us consider the particular case in which the two metallic plates are parallel so that $y_u(x) - y_l(x) = a$ inside the ENZ anisotropic channel, where $a \equiv a_1 = a_2$ (the inset of Fig. 12 represents the particular case in which $y_u(x)$ and $y_l(x)$ vary linearly with x). It is obvious that when $y_u(x) - y_l(x) = a$, we have that $\varphi_u(x) - \varphi_l(x) = 0$. Hence, using (19) it is found that for this very generic family of geometries the magnetic field inside the ENZ channel is given by,

$$H_z(x) = c_1 \cos(k_x x) + c_2 \sin(k_x x), \quad k_x = k_0 \sqrt{\varepsilon_{yy} \mu_{r,p}} \quad (\varphi_u = \varphi_l) \quad (20)$$

where $c_1$ and $c_2$ are unknown coefficients. Next we match the magnetic field $H_z$ and the electric field $E_y$ at the interfaces $x = 0$ and $x = d$. Thus, from (13) and (17) we obtain the following equations with respect to the unknowns $\rho$, $T$, $c_1$ and $c_2$:

$$H_z(0) = H_0^{inc}(1+\rho), \quad H_z(d) = H_0^{inc} T, \quad H'_z(0) = \varepsilon_{yy} i k_0 H_0^{inc}(1-\rho), \quad \text{and}$$

$H'_z(d) = \varepsilon_{yy} i k_0 H_0^{inc} T$, where $H_z(x)$ is given by (20). Solving for the unknowns it is straightforward to obtain that in the $\varepsilon_{xx} = 0$ limit,

$$\rho = \frac{-i(\varepsilon_{yy} - \mu_{r,p}) k_0 d \, \text{sinc}(k_x d)}{2\cos(k_x d) - i(\varepsilon_{yy} + \mu_{r,p}) k_0 d \, \text{sinc}(k_x d)} \quad (21a)$$

$$T = \frac{2}{2\cos(k_x d) - i(\varepsilon_{yy} + \mu_{r,p}) k_0 d \, \text{sinc}(k_x d)} \quad (21b)$$

where $\text{sinc}(x) \equiv \sin(x)/x$. It can be verified that $|T|^2 = 1 - |\rho|^2$ consistently with the conservation of energy theorem. Also, if we let $\varepsilon_{yy} \to 0$ the above formulas reduce to (1), i.e. to the result obtained in our previous work [9] for a channel filled with an isotropic ENZ material. We underline that the above formulas are valid independent of the specific geometry of the ENZ channel, provided $y_u(x) - y_l(x) = a$ (i.e. only the distance between the metallic plates along the *y*-direction must be invariant).

Remarkably, Eq. (21) predicts that when the permittivity of the anisotropic channel along the *y*-direction is chosen such that $\varepsilon_{yy} = \mu_{r,p}$, then the reflection coefficient vanishes, independent of the specific dimensions or parameters of the channel. Thus, an ENZ anisotropic material with such parameters may allow bending the waveguide at will (independent of its length) without affecting the transmission efficiency. Note that the profile $y = y_u(x)$ is completely arbitrary and may have very abrupt or curvy variations. This result confirms that ENZ anisotropic materials may have interesting potentials in improving the transmission characteristic along 'bendy' channels. In order to verify the proposed results, we used CST Microwave Studio™ [14] to calculate the S-parameters of the structure for the particular geometry shown in the inset of Fig. 12 where $y_u(x)$ and $y_l(x)$ vary linearly with $x$. In the simulations we considered that $\varphi = 45°$, $d = 0.77a$, $\mu_{r,p} = 1$, and assumed that $\varepsilon_{xx}$ follows a Drude-type model with plasma frequency $\omega_p$ such that $\frac{\omega_p}{a} c = \frac{\pi}{2}$ and negligible losses ($\Gamma = 0.001 \omega_p$). Using (21) it can be verified that $\rho = 0$ if either $\varepsilon_{yy} = 1$ or if $k_x d = n\pi$ with $n = 1, 2, ...$, which for $n = 1$ corresponds to $\varepsilon_{yy} = 6.3$ (for the considered geometry). These results are completely supported by the full wave simulations depicted in Fig. 12, which confirm, in fact, the validity of (21). We also calculated the reflection characteristic for $\varepsilon_{yy} = 3.4$, and consistently with (21) we obtained that $|\rho| = 0.46$ at the plasma frequency $\omega_p$. Note that the values of $\varepsilon_{yy}$ that guarantee complete transmission are independent of the $\varphi$ angle. As mentioned in section II, the irregular behavior of $\rho$ below $\omega_p$ is due to the excitation of ''quasistatic'' plasmonic resonances. It is also worth mentioning that an ENZ anisotropic material with $\varepsilon_{xx} = 0$ may be synthesized at infrared and optical frequencies by alternately stacking (along the y-direction) slabs of a material with positive permittivity (standard dielectrics) and slabs of a material with negative permittivity (e.g. semiconductor or a noble metal).

## B. Waveguides with Sharp Walls

An interesting geometry that caught our attention is depicted in panel (b) of Fig. 11. It consists of an abrupt transition with thickness $d$ and two 90º-bends. In what follows, we study the effect of filling such waveguide transition with an ENZ anisotropic material. To begin with, it is important to point out that Eq. (21) is not valid for the geometry of Fig. 11b. In fact, (21) was derived under the hypothesis that $y_u(x) - y_l(x) = const.$ and $\varphi_u(x) - \varphi_l(x) = 0$ inside the ENZ channel. Even though, it may be apparent that these conditions are fulfilled for this geometry, in fact, as explained next, they are not. Indeed, we note that at the interfaces $x = 0$ and $x = d$ the walls of the waveguide have sharp transitions with $\varphi_u(0) = \varphi_l(d) = 90º$, and thus at these points the second term of (19) has a singularity. Hence, to obtain the correct limit solution one needs to proceed with special care.

To this end, instead of considering sharp walls with an abrupt transition, we will suppose that the profile of the metallic plates varies continuously in some transition region. To be specific consider the interface $x = 0$ in Fig. 11b, where the upper metallic plate has an abrupt bend. In order to avoid the singularity in the second term of (19), we suppose that the profile of the upper metallic plate, $y_u(x)$, varies linearly from $a_1$ to $a_{ch}$ over some transition region with thickness $\delta$, as illustrated in the inset of Fig. 11b. Notice that the thickness $\delta$ may be arbitrarily small. In what follows we will obtain the solution of (19) in the limit $\delta \to 0^+$.

In the thin transition region $0 < x < \delta$ (filled with ENZ material), (19) can be rewritten as (using $y_l(x) = 0$, $\varphi_l(x) = 0$, and $y_u(x) = a_1 + \tan\varphi_u \, x$ with $\tan(\varphi_u) = \dfrac{a_{ch} - a_1}{\delta}$),

$$\frac{\partial^2 H_z}{\partial x^2} + \frac{\tan(\varphi_u)}{a_1 + \tan(\varphi_u)x} \frac{\partial H_z}{\partial x} + k_x^2 H_z = 0, \qquad \text{for } \varepsilon_{xx} = 0 \tag{22}$$

where $k_x = k_0 \sqrt{\varepsilon_{yy} \mu_{r,p}}$. The general solution of the above equation is:

$$H_z = b_1 J_0\left(k_x\left(x + a_1 \cot \varphi_u\right)\right) + b_2 Y_0\left(k_x\left(x + a_1 \cot \varphi_u\right)\right) \tag{23}$$

where $J_0$ and $Y_0$ are Bessel functions of $1^{st}$ kind and order zero, and $b_1$ and $b_2$ are generic constants. Next we impose the initial boundary conditions $H_z = p$ and $\partial H_z / \partial x = q$ at $x = 0$, where $p$ and $q$ are generic values. Using (23) we can relate the unknown constants $b_1$ and $b_2$ with $p$ and $q$: $b_i = b_i(p, q, \delta)$, $i$=1,2. Finally, we replace $b_i = b_i(p, q, \delta)$, $i$=1,2, in (23), and in this way we obtain $H_z(x) = H_z(x, p, q, \delta)$. In particular, using this procedure $H_z$ and $\frac{\partial H_z}{\partial x}$ can be evaluated at the plane $x = \delta$. In the limit $\delta \to 0^+$ it can be proven that:

$$\lim_{\delta \to 0^+} H_z(x = \delta) = p \tag{24a}$$

$$\lim_{\delta \to 0^+} \frac{\partial H_z}{\partial x}(x = \delta) = \frac{a_1}{a_{ch}} q \tag{24b}$$

Thus, the previous analysis shows that as the wave propagates through the narrow transition region the magnetic field is nearly unchanged, whereas its derivative (proportional to the electric field component $E_y$) varies extremely fast across the very narrow transition. Note that the derived relations guarantee that the flux of power through the interfaces $x = 0$ and $x = \delta$ is invariant. Similar relations are obtained for the interface at $x = d$, in particular we find that $\lim_{\delta \to 0^+} \frac{\partial H_z}{\partial x}(x = d - \delta) = \frac{a_2}{a_{ch}} \frac{\partial H_z}{\partial x}(x = d)$.

We are now ready to calculate the scattering parameters for this propagation problem. As in the previous section, it is clear the magnetic field in the air region is given by (17), whereas the magnetic field in the ENZ region $0^+ < x < d^-$ is given by (20). To compute the reflection and transmission coefficients we need to match the tangential fields at the interfaces. Using (24) to relate the fields inside the ENZ channel (after the transition region) with the fields immediately after the interfaces with air, we obtain the following set of equations: $H_z(0) = H_0^{inc}(1+\rho)$, $H_z(d) = H_0^{inc}T$, $H_z'(0^+) = \varepsilon_{yy} i k_0 H_0^{inc}(1-\rho)a_1/a_{ch}$, and $H_z'(d^-) = \varepsilon_{yy} i k_0 H_0^{inc} T a_2/a_{ch}$, where $H_z(x)$ is given by (20). Solving this system with respect to unknowns $\rho$, $T$, $c_1$ and $c_2$, we obtain the following expressions for the reflection and transmission coefficients (valid in $\varepsilon_{xx} = 0$ limit):

$$\rho = \frac{(a_1 - a_2)\cos(k_x d) - i\left(\varepsilon_{yy}\frac{a_1 a_2}{a_{ch}} - \mu_{r,p} a_{ch}\right) k_0 d\ \text{sinc}(k_x d)}{(a_1 + a_2)\cos(k_x d) - i\left(\varepsilon_{yy}\frac{a_1 a_2}{a_{ch}} + \mu_{r,p} a_{ch}\right) k_0 d\ \text{sinc}(k_x d)} \quad (25a)$$

$$T = \frac{2a_1}{(a_1 + a_2)\cos(k_x d) - i\left(\varepsilon_{yy}\frac{a_1 a_2}{a_{ch}} + \mu_{r,p} a_{ch}\right) k_0 d\ \text{sinc}(k_x d)} \quad (25b)$$

It can be verified that $1 = |\rho|^2 + \frac{a_2}{a_1}|T|^2$, consistently with the conservation of the power flow. Considering the important case $a \equiv a_1 = a_2$ the above formulas demonstrate that in the $\varepsilon_{xx} = 0$ limit the wave may tunnel through the ENZ transition if either $\varepsilon_{yy} = \mu_{r,p}\frac{a_{ch}^2}{a_1 a_2}$ or if $k_x d = n\pi$, (n=1,2,...). Both these conditions yield $\rho = 0$ at the considered frequency, and correspond to geometrical resonances of the structure (i.e. depend on the specific geometrical parameters of the ENZ transition). However, a simple analysis shows that the reflection coefficient may also be made negligibly small provided $\varepsilon_{yy} a^2 / a_{ch}^2 \ll \mu_{r,p}$ and $k_0 \mu_{r,p} A_p / a \ll 1$ where $A_p = a_{ch} d$ is the area of the plasmonic channel. The latter condition is precisely the one derived in our previous work [9] for propagation across isotropic ENZ channels, while the former condition is easily satisfied for standard material values ($\varepsilon_{yy} \sim 1$, $\mu_{r,p} = 1$) and for long channels ($a_{ch} \gg a$). Thus, we conclude that for the geometry of panel b) of Fig. 11 and for typical material parameters, the transmissivity of a narrow long channel is approximately independent of the channel being filled with an isotropic ENZ material or an anisotropic ENZ material (with $\varepsilon_{xx} = 0$). In particular, it may be possible to squeeze more and more energy through the narrow channel provided the thickness $d$ is made increasingly small. In this regime, the ENZ channel supports a zero-order resonance, consistently with the discussion of section II.A. This result confirms that, in fact, in some propagation scenarios ENZ anisotropic materials may have potentials similar to those of ENZ isotropic materials. (The results of [9] also support such conclusions for a geometry analogous to that of Fig. 11b, but with an 180º bend. The formula for the reflection coefficient presented in [9] can be derived using arguments similar to those employed here).

In order to illustrate some of the suggested potentials, we have used CST Microwave Studio[TM] [14] to characterize a very narrow waveguide transition (Fig. 11b) with $a \equiv a_1 = a_2$, $d = 0.05a$, and $a_{ch} = 8a$. The plasma frequency is chosen such that $\omega_p a / c = 0.4\pi$. In Fig. 13 we plot the corresponding transmission (S21) parameter supposing that the channel is filled either with an anisotropic ENZ material with $\varepsilon_{xx} = 0$ (at the design frequency) and $\varepsilon_{yy} = 1.0$ (solid red line), or with an isotropic ENZ material (dashed red line). For simplicity, the losses are assumed negligible. Consistently, with our theoretical analysis it is found that the transmitted power is nearly independent of the material being isotropic or anisotropic at $\omega \approx \omega_p$. Moreover, it is seen that transmission is greatly enhanced as compared to the case in which the transition is filled with air (black line). It is important to note that if the channel is filled with an anisotropic ENZ material with parameters such that $\varepsilon_{xx} = 1$ and $\varepsilon_{yy} = 0$, then no tunneling effect is observed (not shown here) and the transmission level is similar to that of the empty channel.

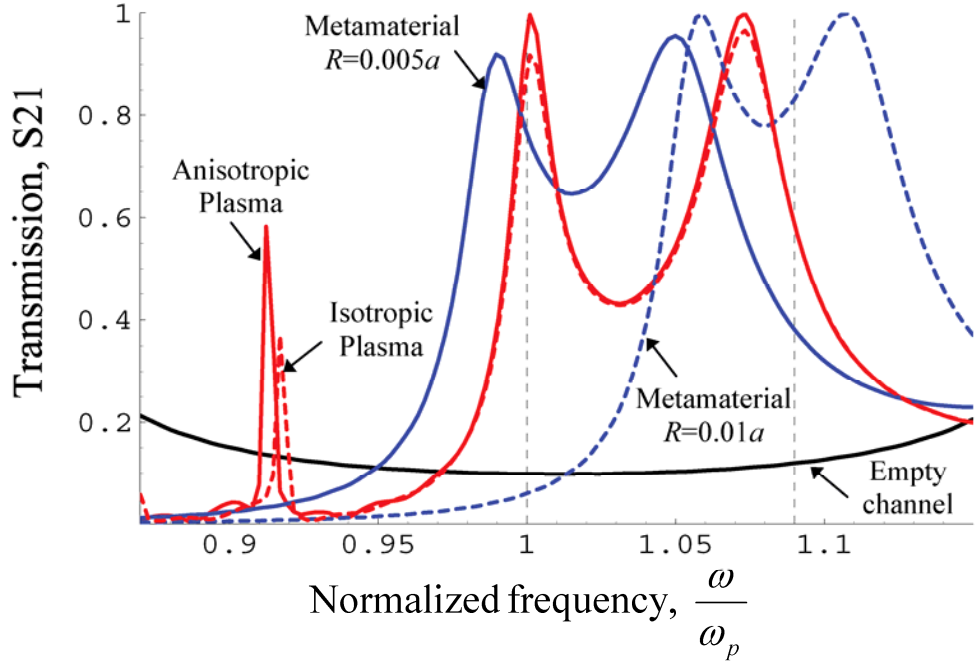

**Fig. 13**. (Color online) Amplitude of the transmission coefficient (S21-parameter) as a function of normalized frequency for a waveguide with sharp walls (Fig. 11b). The parameters of the waveguide are $a \equiv a_1 = a_2$, $d = 0.05a$, and $a_{ch} = 8a$. (i) Black line: Empty transition. (ii) Red line (light gray in grayscale): the transition is filled either with an anisotropic ENZ material (solid line) or with an isotropic ENZ material (dashed line). (iii) Blue line (dark gray in grayscale): waveguide is filled with metallic wires (wire medium metamaterial) with radius $R = 0.005a$ (solid line) or $R = 0.01a$ (dashed line).

As discussed in the beginning of this section, ENZ anisotropic materials may be synthesized as metamaterials. As in [9], next we explore the possibility of using wire media [22]-[23] to simulate a continuous ENZ anisotropic material. To this end, we consider that an array of PEC wires is placed inside the ENZ transition. The wires stand in air and are centered at the points $(0, 0.5+m, n)a$, where $m = 0, 1, ..., 7$, and $n = 0, \pm 1, \pm 2, ...$. Note that the spacing between the wires is $a$, i.e. the lattice constant is equal to distance between the parallel-plates in the air regions. The wires are directed along the $x$-direction ($0 < x < d$), have length $d$ and radius $R$. Note that the wires corresponding to the indices $m = 1, 2, ...6$ have both extremities connected to the metallic plates (the wires corresponding to indices $m = 0$ and $m = 7$ have only one extremity connected to a metallic plate). The radius of the wires is calculated so that the corresponding metamaterial plasma frequency also satisfies $\omega_p a / c = 0.4\pi$. From [22] we know that to a first-order approximation $\left(\dfrac{\omega_p}{c}a\right)^2 = \dfrac{2\pi}{\ln\left(\dfrac{a}{2\pi\, r_w}\right) + 0.5275}$, and thus we find that the required radius is $R = 0.005a$. The S21-parameter dispersion calculated for such metamaterial is depicted in Fig. 13 (solid blue line). It is seen that as predicted by our theory, the transmission is significantly enhanced near $\omega = \omega_p$, even though there is a small downshift in frequency as compared to the design specification. It is important to note that even though the electrical length of the wires is very small, the array of wires emulates indeed an ENZ metamaterial slab. The reason is that (with the exception of the wires associated with the indices $m = 0$ and $m = 7$) the wires are physically connected across the metallic plates, and this increases their effective electrical length. Thus, at least in the region $a < y < a_{ch} - a$ the wire medium behaves indeed as an ENZ anisotropic material. We have also studied the effect of varying the radius of the wires. For $R = 0.01a$ it is expected that the plasma frequency of the metamaterial increases about 9%: $\omega'_p = 1.09\omega_p$ (dashed vertical line in Fig. 13). The transmission characteristic for a channel filled with such metamaterial is shown in Fig. 13 (dashed blue line). As seen, the peak of transmission is moved to higher frequencies, consistently to what one would expect from the previous analysis. As in the case $R = 0.005a$, for $R = 0.01a$ the peak of transmission is also slightly downshifted with respect to the design frequency. The reason for this property may be related with the discreteness of the metamaterial design (note that we use only 8 rows of wires), and probably also to the fact that the wires corresponding to indices $m = 0$ and $m = 7$ are not able to emulate the ENZ property since one of their extremities is unconnected.

In conclusion, we have demonstrated theoretically and numerically that ENZ anisotropic materials may have common features with ENZ isotropic materials in some propagation scenarios. The condition $\varepsilon_{xx} = 0$ effectively freezes the wavefronts inside the anisotropic ENZ material along the *y*-direction (i.e. the wavefronts are normal to the *x*-direction). It was discussed that for some geometries the transmission along anisotropic ENZ channels may be independent of the specific geometry of channel, and that a zero-order resonance may enable a tunneling effect similar to that reported in [9]. In addition, we have further discussed the possible realization of ENZ metamaterials using wire media, and enlightened some key features of such metamaterial implementations.

## IV. Conclusions

In this work, we have further investigated the applications and properties of the supercoupling, field confinement, and tunneling phenomenon identified in [9]. It was shown that waveguide channels filled with ENZ materials support a zero-order resonance, which may enable anomalous transmission and supercoupling of energy through very narrow channels or abrupt bends, independent of their specific geometry. We demonstrated that the root of the tunneling effect is the property that the current injected in a metallic plate remains constant inside the ENZ channel. It was analytically proved that the effect of metallic losses on the waveguide walls may be of second order, and thus that in principle the dielectric loss in the ENZ material is the dominant loss mechanism. It was suggested that ENZ materials may enable electric field concentration and confinement in a subwavelength air cavity. Two different realistic 3D configurations based on the concept of simulated plasma were proposed to emulate the studied propagation scenarios at microwaves. In addition, we developed a theory to describe the propagation of electromagnetic waves through anisotropic ENZ materials. It was proved that in some very generic propagation scenarios the scattering coefficients can be calculated in closed analytical form. Our analysis demonstrates that in some circumstances the scattering coefficients may be made independent of the specific geometry of channel. Furthermore, in some cases, ENZ anisotropic channels may also enable the anomalous transmission of energy through very narrow metallic channels, independent of the specific dimensions of the channel (zero-order resonance). Finally, we discussed some key properties of anisotropic ENZ metamaterials made of an array of microstructured wires.

**Acknowledgments:**

This work is supported in part by the U.S. Office of Naval Research (ONR) grant number N 00014 -07-1-0622. Mário Silveirinha has been partially supported by a fellowship from "Fundação para a Ciência e a Tecnologia".

# Appendix A

Here, we discuss the singularities of electromagnetic fields near a junction between several dielectric and metallic materials [24, pp. 23]. The novelty of our analysis is that we study the singularities at a sharp corner involving materials with positive permittivity and materials with negative permittivity. Our results establish that near an edge the fields may have very strong singularities, and that in the lossless limit, the fields may not be square integrable.

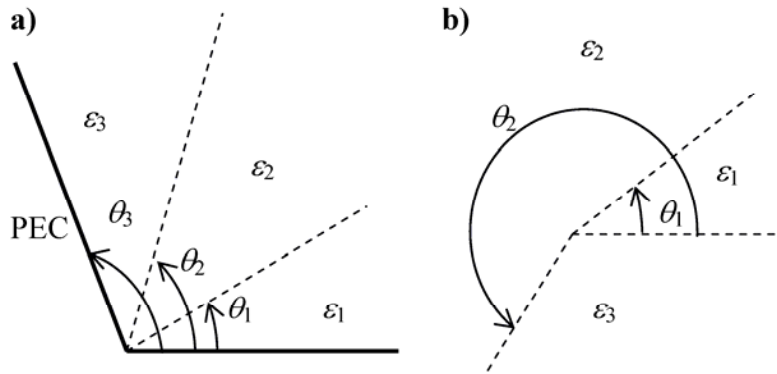

**Fig. 14**. Panel (a): Corner between a PEC region and three-dielectric wedges. Panel (b): Corner between three dielectric wedges.

To begin with, we consider the geometry of panel (a) of Fig. 14, which depicts a corner between a PEC region and three dielectric wedges (uniform along $z$). As in the previous sections, we suppose that the fields are H-polarized, with $\mathbf{H} = H_z(x, y)\hat{\mathbf{u}}_z$. The magnetic field satisfies,

$$\frac{1}{\varepsilon(\theta)}\frac{1}{r}\frac{\partial}{\partial r}\left(r\frac{\partial H_z}{\partial r}\right) + \frac{1}{r^2}\frac{\partial}{\partial \theta}\frac{1}{\varepsilon(\theta)}\frac{\partial}{\partial \theta}H_z + k_0^2 H_z = 0 \tag{A1}$$

where $(r, \theta)$ are associated with the system of polar coordinates centered at the junction between the materials, and $\varepsilon = \varepsilon(\theta)$ is the sectionally constant permittivity near the corner. We are interested in the quasi-static regime, where the third term ($k_0^2 H_z$) in the left-hand side of (A1) can be neglected. In that case, (A1) becomes separable and admits solutions of the form,

$$H_\lambda = r^\lambda f_\lambda(\theta) \quad ; \quad H_{-\lambda} = r^{-\lambda} f_\lambda(\theta) \tag{A2}$$

(for $\lambda = 0$, $H_{-\lambda}$ should be defined as $H_{-\lambda} = \ln r \, f_\lambda(\theta)$) where $-\lambda^2$ is an eigenvalue ($\lambda$ is defined in such a way that $\text{Re}\{\lambda\} \geq 0$) and $f_\lambda$ is the corresponding eigenfunction of the integral operator:

$$\mathbf{L} = \varepsilon(\theta)\frac{\partial}{\partial \theta}\frac{1}{\varepsilon(\theta)}\frac{\partial}{\partial \theta} \tag{A3}$$

Thus, we have that $\mathbf{L}f_\lambda = -\lambda^2 f_\lambda$. It is clear from (A2) that the behavior of the fields near the corner is determined by the eigenvalues of $\mathbf{L}$. Next we determine the characteristic equation for the eigenvalues. For the geometry of Fig. 14 it is simple to verify that (imposing $f'(\theta) = 0$ at $\theta = 0$ and $\theta = \theta_3$, so that the tangential electric field vanishes at the PEC plates),

$$f_\lambda(\theta) = \begin{cases} c_1 \cos(\lambda\theta) & 0 < \theta < \theta_1 \\ c_2 \cos(\lambda(\theta-\theta_1)) + c_3 \sin(\lambda(\theta-\theta_1)) & \theta_1 < \theta < \theta_2 \\ c_4 \cos(\lambda(\theta-\theta_3)) & \theta_2 < \theta < \theta_3 \end{cases} \quad \text{(A4)}$$

where $c_1, \ldots c_4$ are unknown coefficients. At the dielectric interfaces it is necessary that the tangential fields are continuous, and hence $[f] = 0$ and $[f'/\varepsilon] = 0$, where $[\ldots]$ represents the jump discontinuity of the quantity inside brackets at the interface. Imposing the conditions $[f] = 0$ and $[f'/\varepsilon] = 0$ at $\theta = \theta_1$ and $\theta = \theta_2$, we obtain an homogenous 4×4 linear system for the unknown coefficients. A non-trivial solution can exist only if the determinant of the corresponding matrix vanishes. In this way, we find that the eigenvalues must satisfy the following characteristic equation:

$$\lambda^2 \left( \frac{\sin(\lambda\Delta_1)\cos(\lambda\Delta_2)\cos(\lambda\Delta_3)}{\varepsilon_1\varepsilon_2} + \frac{\cos(\lambda\Delta_1)\cos(\lambda\Delta_2)\sin(\lambda\Delta_3)}{\varepsilon_2\varepsilon_3} + \right.$$
$$\left. + \frac{\cos(\lambda\Delta_1)\sin(\lambda\Delta_2)\cos(\lambda\Delta_3)}{\varepsilon_2^2} - \frac{\sin(\lambda\Delta_1)\sin(\lambda\Delta_2)\sin(\lambda\Delta_3)}{\varepsilon_1\varepsilon_3} \right) = 0 \quad \text{(A5)}$$

where we put $\Delta_1 = \theta_1 - 0$, $\Delta_2 = \theta_2 - \theta_1$, and $\Delta_3 = \theta_3 - \theta_2$. Note that $\lambda = 0$ is always an eigenvalue, independent of the considered materials. In general, the remaining eigenvalues have to be computed using numerical methods.

Let us consider first that the materials have some losses so that the corresponding $\varepsilon$ has a small positive imaginary component. In order that the solutions yielded by (A2) have physical meaning it is necessary that the corresponding electromagnetic fields are square-integrable near the corner (this condition is necessary to ensure that the power dissipated in the materials is finite), i.e. $\int |H_z|^2 r\,dr\,d\theta < \infty$ and $\int |\mathbf{E}|^2 r\,dr\,d\theta < \infty$. The critical components of the fields are $E_\theta$ and $E_r$. It is simple to verify that the electric field associated with $H_\lambda = r^\lambda f_\lambda(\theta)$ is square-integrable only if $\text{Re}\{\lambda\} > 0$ or $\lambda = 0$, while the field associated with $H_{-\lambda}$ is square-integrable only if $\text{Re}\{\lambda\} < 0$. Thus, it is clear that only one of the solutions defined by (A2) may have physical meaning (since we admit $\text{Re}\{\lambda\} \geq 0$, that solution is necessarily $H_\lambda$).

We also note that by integrating the identity $\dfrac{d}{d\theta}\left(f^* \dfrac{1}{\varepsilon}\dfrac{df}{d\theta}\right) = \dfrac{1}{\varepsilon}\left(\left|\dfrac{df}{d\theta}\right|^2 + f^*\mathbf{L}f\right)$ over the interval $[0, \theta_3]$, we find that the eigenfunctions of $\mathbf{L}$ satisfy:

$$\int_0^{\theta_3} \frac{1}{\varepsilon(\theta)}\left(\left|\frac{df_\lambda}{d\theta}\right|^2 - \lambda^2 |f_\lambda|^2\right)d\theta = 0 \tag{A6}$$

In particular, the previous equation clearly shows that if the materials have losses ($\text{Im}\{\varepsilon\} > 0$) the operator $\mathbf{L}$ cannot have an eigenvalue such that $\text{Re}\{\lambda\} = 0$ and $\lambda \neq 0$, i.e. provided the materials have some losses, $\lambda$ cannot be purely imaginary (or equivalently $\lambda^2$ cannot be a negative real number). This means that in the lossy case ($\text{Im}\{\varepsilon\} > 0$) the eigenfunction $H_\lambda = r^\lambda f_\lambda(\theta)$ is always a physical solution of the problem, and the corresponding electromagnetic fields are square-integrable near the corner.

But what happens in the lossless limit? Are still all the eigenvalues such that $\lambda^2$ cannot be a negative real number? If either $\varepsilon(\theta) > 0$ (i.e. all the materials in Fig. 14a have positive dielectric constant) or $\varepsilon(\theta) < 0$ (all the materials in Fig. 14a have negative dielectric constant), it is evident from (A6) that $\lambda^2$ cannot be real negative for a non-trivial eigenfunction $f_\lambda$. Thus, when the permittivity of all materials has the same sign, we conclude that the electromagnetic fields are still square-integrable near the corner.

However, as proved next, if the sign of the permittivity of the materials in different dielectric regions is not the same, then the situation changes drastically.

Let us consider for example that $\varepsilon_3 = \varepsilon_1$ and $\Delta_3 = \Delta_1$ in Fig. 14a. In that case it can be verified that the $\lambda \neq 0$ solutions of the characteristic equation (A5) are such that:

$$\frac{\varepsilon_2}{\varepsilon_1} = -\cot(\lambda\Delta_1)\tan\left(\frac{\lambda\Delta_2}{2}\right) \quad \text{or} \quad \frac{\varepsilon_2}{\varepsilon_1} = +\cot(\lambda\Delta_1)\cot\left(\frac{\lambda\Delta_2}{2}\right) \tag{A7}$$

Thus, in these circumstances, the eigenvalues are functions $\lambda = \lambda(\varepsilon_{2,1})$ of the dielectric contrast $\varepsilon_{2,1} = \varepsilon_2/\varepsilon_1$. Let us find for which values of $\varepsilon_{2,1}$ may the characteristic equation admit purely imaginary solutions for $\lambda$, i.e. solutions of the form $\lambda = i\xi$, with $\xi$ a real number. Substituting $\lambda = i\xi$ in (A7), it can be easily verified that for $-\infty < \xi < \infty$ we obtain that:

$$-\infty < \frac{\varepsilon_2}{\varepsilon_1} < \max\left\{-1, -\frac{\Delta_2}{2\Delta_1}\right\} \tag{A8}$$

Hence, we conclude that when materials are lossless and the permittivity contrast is such that the above condition is verified, then operator $\mathbf{L}$ has an eigenvalue such that $\lambda$ is purely imaginary. This means that in these conditions the electromagnetic fields have a strong singularity near the corner and in particular the fields are not square-integrable. For example, for the geometry of Fig. 3 (waveguide with a 90º-bend filled with a plasmonic material), we have $\Delta_1 = \Delta_2 = 90[\deg]$ and $\varepsilon_1 = 1$. Hence, (A8) predicts that for such geometry the electromagnetic fields may have very strong singularities near the corner when the permittivity of the plasmonic material is such that $\varepsilon_2 < -1/2$. This justifies the resonant behavior of the transmission characteristic of the waveguide for frequencies below the plasma frequency. Note that even though for small losses the fields are square-integrable, the singularities can still be very significant, because some eigenvalues are such that $\text{Re}\{\lambda\} \approx 0$ with $\text{Im}\{\lambda\}$ very large.

For the sake of completeness, next we briefly discuss the singularities of the electromagnetic fields near a junction between three dielectric materials (panel (b) of Fig. 14). Proceeding as in the previous case, we may easily find that the characteristic equation for the eigenvalues of $\mathbf{L}$ is now:

$$\lambda^3 \left( \left( \frac{1}{\varepsilon_2^2 \varepsilon_1} + \frac{1}{\varepsilon_3^2 \varepsilon_1} \right) \cos(\lambda\Delta_1)\sin(\lambda\Delta_2)\sin(\lambda\Delta_3) + \left( \frac{1}{\varepsilon_1^2 \varepsilon_2} + \frac{1}{\varepsilon_3^2 \varepsilon_2} \right) \sin(\lambda\Delta_1)\cos(\lambda\Delta_2)\sin(\lambda\Delta_3) + \right.$$
$$\left. + \left( \frac{1}{\varepsilon_1^2 \varepsilon_3} + \frac{1}{\varepsilon_2^2 \varepsilon_3} \right) \sin(\lambda\Delta_1)\sin(\lambda\Delta_2)\cos(\lambda\Delta_3) + \frac{2(1 - \cos(\lambda\Delta_1)\cos(\lambda\Delta_2)\cos(\lambda\Delta_3))}{\varepsilon_1 \varepsilon_2 \varepsilon_3} \right) = 0$$

(A9)

where we put $\Delta_1 = \theta_1 - 0$, $\Delta_2 = \theta_2 - \theta_1$, and $\Delta_3 = 2\pi - \theta_2$. It can be verified that in the particular case $\Delta_3 = 0$ (i.e. when the junction consists of only two materials), the eigenvalues of $\mathbf{L}$ may be such that $\lambda$ is purely imaginary when:

$$\frac{\varepsilon_2}{\varepsilon_1} \in \text{Interval}\left\{ -\frac{\Delta_1}{\Delta_2}, -\frac{\Delta_2}{\Delta_1} \right\} \qquad \text{(for } \Delta_3 = 0\text{)} \tag{A10}$$

For example, for $\Delta_1 = \Delta_2 = 180°$ the above equation predicts that the fields may have a strong singularity when $\varepsilon_2 / \varepsilon_1 = -1$, which is the well-known resonance condition for a planar interface between two dielectric media.